%% file: clean_version_after_referee_comments.tex
\documentclass[longauth]{aa}
\input{mycommands}

\usepackage{graphicx}	
\usepackage{amsmath}	
\usepackage{multirow}
\usepackage{siunitx}
\usepackage{lipsum}
\usepackage{pifont}
\usepackage{xcolor}
\usepackage{txfonts}
\usepackage{bm}	        
\usepackage{siunitx} 
\usepackage[utf8]{inputenc}
\usepackage{xparse}
\usepackage[pdfpagelabels=false]{hyperref}	
\hypersetup{colorlinks=true,linkcolor=blue,citecolor=blue,filecolor=blue,urlcolor=blue,}

\usepackage{adjustbox} 
\usepackage{CJKutf8}

\usepackage{svg}
\newcommand{\orcid}[1]{\href{https://orcid.org/#1}{\includegraphics[width=10pt]{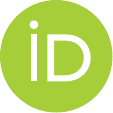}}}

\newcommand{\Bonn}{Argelander-Institut f\"ur Astronomie, Universit\"at Bonn, Auf dem H\"ugel 71, 53121 Bonn, Germany}

\newcommand{\NRAO}{National Radio Astronomy Observatory, 520 Edgemont Road, Charlottesville, VA 22903, USA}

\newcommand{\OSU}{Department of Astronomy, The Ohio State University, 140 West 18th Ave, Columbus, OH 43210, USA}

\newcommand{\CCAP}{Center for Cosmology and Astroparticle Physics, 191 West Woodruff Avenue, Columbus, OH 43210, USA}

\newcommand{\Princeton}{Department of Astrophysical Sciences, Princeton University, 4 Ivy Lane, Princeton, NJ 08544, USA}

\newcommand{\MPIA}{Max-Planck-Institut f\"{u}r Astronomie, K\"{o}nigstuhl 17, D-69117, Heidelberg, Germany}

\newcommand{\HeidelbergARI}{Astronomisches Rechen-Institut, Zentrum f\"{u}r Astronomie der Universit\"{a}t Heidelberg, M\"{o}nchhofstra\ss e 12-14, D-69120 Heidelberg, Germany}

\newcommand{\HeidelbergITA}{Universit\"{a}t Heidelberg, Zentrum f\"{u}r Astronomie, Institut f\"{u}r Theoretische Astrophysik, Albert-Ueberle-Stra{\ss}e 2,69120 Heidelberg, Germany}

\newcommand{\ESO}{European Southern Observatory, Karl-Schwarzschild Stra{\ss}e 2, D-85748 Garching bei M\"{u}nchen, Germany}

\newcommand{\CfA}{Center for Astrophysics $\mid$ Harvard \& Smithsonian, 60 Garden Street, Cambridge, MA 02138, USA}

\newcommand{\Alberta}{Department of Physics, University of Alberta, Edmonton, AB T6G 2E1, Canada}

\newcommand{\OAN}{Observatorio Astron{\'o}mico Nacional (IGN), C/ Alfonso XII, 3, E-28014 Madrid, Spain}

\newcommand{\astron}{Netherlands Institute for Radio Astronomy (ASTRON),  Oude Hoogeveensedijk 4, 7991 PD Dwingeloo, Netherlands}

\newcommand{\kapeyn}{Kapteyn Astronomical Institute, University of Groningen, PO Box 800, 9700 AV Groningen, The Netherlands}

\newcommand{\UCT}{Department of Astronomy, University of Cape Town, Private Bag X3, 7701 Rondebosch, South Africa}

\newcommand{\MPE}{Max-Planck-Institut f\"{u}r extraterrestrische Physik, Giessenbachstra{\ss}e 1, D-85748 Garching, Germany}

\newcommand{\Conneticut}{Department of Physics, University of Connecticut, Storrs, CT, 06269, USA}

\newcommand{\ASIAA}{Institute of Astronomy and Astrophysics, Academia Sinica, Astronomy-Mathematics Building, No. 1, Sec. 4, Roosevelt Road, Taipei 10617, Taiwan}

\newcommand{\NRAOsoc}{National Radio Astronomy Observatory, 1003 Lopezville Road, Socorro, NM 87801, USA}

\newcommand{\Wyoming}{Department of Physics \& Astronomy, University of Wyoming, Laramie, WY 82071}

\newcommand{\Tamkang}{Department of Physics, Tamkang University, No.151, Yingzhuan Road, Tamsui District, New Taipei City 251301, Taiwan}

\newcommand{\SanDiego}{Center for Astrophysics and Space Sciences, Department of Physics, University of California San Diego, 9500 Gilman Drive, La Jolla, CA 92093, USA}

\newcommand{\Oxford}{Sub-department of Astrophysics, Department of Physics, University of Oxford, Keble Road, Oxford OX1 3RH, UK}

\newcommand{\COOL}{Cosmic Origins Of Life (COOL) Research DAO, \href{https://coolresearch.io}{https://coolresearch.io}}

\begin{document} 

    \title{PHANGS-MeerKAT and MHONGOOSE HI observations of nearby spiral galaxies: physical drivers of the molecular gas fraction, $R_{\mathrm{mol}}$}

    \titlerunning{MeerKAT HI observations of nearby spiral galaxies: the physical drivers of $R_{\mathrm{mol}}$}

   \author{
   Cosima~Eibensteiner\inst{1},\inst{2}\fnmsep\thanks{\email{ceibenst@nrao.edu}} 
   \thanks{Jansky Fellow of the National Radio Astronomy Observatory}\orcid{0000-0002-1185-2810}
   \and 
   Jiayi~Sun \begin{CJK*}{UTF8}{gbsn}(孙嘉懿)\end{CJK*}\inst{3}\fnmsep\thanks{NASA Hubble Fellow}\orcid{0000-0003-0378-4667} 
   \and
   Frank Bigiel\inst{1}\orcid{0000-0003-0166-9745}
   \and
   Adam~K.~Leroy\inst{4,5}\orcid{0000-0002-2545-1700}
   \and
   Eva~Schinnerer\inst{6}\orcid{0000-0002-3933-7677}
   \and
   Erik~Rosolowsky\inst{7}\orcid{0000-0002-5204-2259}
   \and
   Sushma~Kurapati\inst{8}\orcid{0000-0001-6615-5492}
   \and
   D.~J.~Pisano\inst{8}\orcid{0000-0001-7996-7860}
   \and
   W.~J.~G~de Blok\inst{8,9,10}\orcid{0000-0001-8957-4518}
   \and
   Ashley~T.~Barnes\inst{11}\orcid{0000-0003-0410-4504}
   \and
   Mallory~Thorp\inst{1}\orcid{0000-0003-0080-8547}
   \and
   Dario~Colombo\inst{1}\orcid{0000-0001-6498-2945}
   \and
   Eric~W.~Koch\inst{12}\orcid{0000-0001-9605-780X}
   \and
   I-Da Chiang (\begin{CJK*}{UTF8}{bkai}江宜達 \end{CJK*})
     \inst{13}\orcid{0000-0003-2551-7148}
   \and
   Eve~C.~Ostriker\inst{3}\orcid{0000-0002-0509-9113}
   \and
   Eric~J.~Murphy\inst{2}\orcid{0000-0001-7089-7325}
   \and
   Nikki~Zabel\inst{8}\orcid{0000-0001-7732-5338}
   \and
   Sebstian~Laudage\inst{1}
   \and
   Filippo~M.~Maccagni\inst{10}\orcid{0000-0002-9930-1844}
   \and 
   Julia~Healy\inst{10}\orcid{0000-0003-1020-8684}
   \and
   Srikrishna~Sekhar\inst{14}\orcid{0000-0002-8418-9001}
   \and 
   Dyas~Utomo\inst{2}\orcid{0000-0003-4161-2639}
   \and
   Jakob~den~Brok\inst{10}\orcid{0000-0002-8760-6157}
   \and
   Yixian~Cao\inst{15}\orcid{0000-0001-5301-1326}
   \and
   M\'elanie~Chevance\inst{16,17}\orcid{0000-0002-5635-5180}
   \and
   Daniel~A.~Dale\inst{18}\orcid{0000-0002-5782-9093}
   \and
   Christopher M. Faesi\inst{19}\orcid{0000-0001-5310-467X}
   \and
   Simon~C.~O.~Glover\inst{16}\orcid{0000-0001-6708-1317}
   \and
   Hao~He \begin{CJK*}{UTF8}{gbsn}(何浩)\end{CJK*}\inst{20}\orcid{0000-0001-9020-1858}
   \and
   Sarah~Jeffreson\inst{12}
   \and
   Mar\'ia~J.~Jim\'enez-Donaire\inst{21}\orcid{0000-0002-9165-8080}
   \and
   Ralf~Klessen\inst{22}\orcid{0000-0002-0560-3172}
   \and
   Justus~Neumann\inst{6}\orcid{0000-0002-3289-8914}
   \and
   Hsi-An~Pan\inst{23}\orcid{0000-0002-1370-6964}
   \and
   Debosmita~Pathak\inst{4}
   \and
   Miguel~Querejeta\inst{21}\orcid{0000-0002-0472-1011}
   \and
   Yu-Hsuan Teng\inst{24}\orcid{0000-0003-4209-1599}
   \and
   Antonio~Usero\inst{21}\orcid{0000-0003-1242-505X}
   \and
   Thomas G. Williams\inst{25}\orcid{0000-0002-0012-2142}
   }
   \institute{\Bonn\
              \and
              \NRAO\ 
              \and
              \Princeton
              \and
              \OSU
              \and
              \CCAP
              \and
              \MPIA
              \and
              \Alberta
              \and
              \UCT
              \and
              \kapeyn
              \and
              \astron
              \and
              \ESO
              \and
              \CfA
              \and
              \ASIAA
              \and
              \NRAOsoc
              \and
              \MPE
              \and
              \HeidelbergITA
              \and
              \COOL
              \and
              \Wyoming
              \and
              \Conneticut
              \and
              Hao~He Affiliation
              \and
              \OAN
              \and
              \HeidelbergARI
              \and
              \Tamkang
              \and
              \SanDiego
              \and
              \Oxford
              }

  \date{Received 11 March 2024 ; accepted 1 July 2024}

 \abstract{The molecular-to-atomic gas ratio is crucial to our understanding of the evolution of the interstellar medium (ISM) in galaxies. We investigate the balance between the atomic (\sigatom) and molecular gas (\sigmol) surface densities in eight nearby star-forming galaxies using new high-quality observations from MeerKAT and ALMA (for \hi\ and CO, respectively). We define the molecular gas ratio as $R_{\rm mol} = \sigmol / \sigatom$ and measure how \rmol\ depends on local conditions in the galaxy disks using multi-wavelength observations. We find that, depending on the galaxy, \hi\ is detected at ${>}3\sigma$ out to $20-120$\,kpc in galactocentric radius ($r_{\rm gal}$). The typical radius at which \sigatom\ reaches 1~\uSig\ is  $r_{\hi}\approx22$~kpc, which corresponds to $1-3$ times the optical radius (r$_{25}$). $R_{\rm mol}$ correlates best with the dynamical equilibrium pressure, P$_{\rm DE}$, among potential drivers studied, with a median correlation coefficient of $\left<\rho\right>=0.89$. Correlations between $R_{\rm mol}$ and star formation rate surface density, total gas surface density, stellar surface density, metallicity, and \sigsfr/\Pdyn\ (a proxy for the combined effect of UV radiation field and number density) are present but somewhat weaker. Our results also show a direct correlation between \Pdyn\ and $\Sigma_{\rm SFR}$, supporting self-regulation models. Quantitatively, we measure similar scalings as previous works, and attribute the modest differences that we do find to the effect of varying resolution and sensitivity. At $r_{\rm gal} {\gtrsim}0.4~r_{25}$, atomic gas dominates over molecular gas among our studied galaxies, and at the balance of these two gas phases ($R_{\rm mol}$ = 1), we find that the baryon mass is dominated by stars, with \sigstar > 5~\siggas. Our study constitutes an important step in the statistical investigation of how local galaxy properties (stellar mass, star formation rate, or morphology) impact the conversion from atomic to molecular gas in nearby galaxies.}


   \maketitle
%

\renewcommand{\equationautorefname}{Eq.}
\renewcommand{\sectionautorefname}{Section}
\renewcommand{\subsectionautorefname}{Section}
\renewcommand{\subsubsectionautorefname}{Section}

\section{Introduction}

%
%
%
%

 The accretion of cold gas from the circumgalactic medium is an important driver in the formation and evolution of galaxies. This cold gas first forms atomic hydrogen (\hi) and then molecular hydrogen (\htwo), some of which eventually collapses under gravitational instability and forms new stars \citep[][]{MacLow2004,McKee2007,Klessen2016,Girichidis2020,Chevance2023}. Stellar Feedback also plays an important role in regulating the cold gas phase balance in galaxies via photodissociation of \htwo\ into \hi\ due to radiation emitted by young stars. Consequently, \hi\ is not only an intermediate gas phase in the star formation process but also one of its products \citep[e.g.][]{Allen1997,Heiner2011}. The phase balance between atomic and molecular gas, often parameterized as the atomic-to-molecular gas ratio $R_{\rm mol} = \sigmol/\sigatom$, encodes important information as to the evolution of the interstellar medium (ISM) in galaxies, from local environments in the Milky Way to distant, cold gas reservoirs in high redshift systems.

Atomic gas reservoirs are massive, extended, and often surround spiral galaxies far beyond their optical disks out to  2-5 $\times$ the optical radius, $r_{25}$~\citep[e.g.,][]{Tamburro2009,Wang2016,Eibensteiner2023}. The majority of the gas mass in galaxy disks is in \hi\, but star formation occurs in \htwo\ gas \citep[see e.g.][]{Schruba2011}. The current abundance of molecular gas in galaxies reflects a balance between molecular cloud formation and destruction, with both large-scale dynamical processes and small-scale ISM physics playing important roles \citep[e.g. see][]{Dobbs2014,Sternberg2014,Saintonge2022,Chevance2023}. On large scales, both theory and observations show that the mean pressure or density of the ISM affects the molecular-to-atomic gas ratio \citep[e.g.,][]{Elmegreen1989,Wong2002,Blitz2004,Blitz2006,Leroy2008,Ostriker2010,MacLow2012,Sun2020_pde} as do large-scale dynamical features such as spiral arm-induced shocks \citep{Dobbs2014}. Locally, metallicity and gas column density set the amount of shielding \hi\ provides, which if sufficient can allow molecular gas to become dominant \citep[][]{Krumholz2010,Wolfire2010,Glover2012,Sternberg2014,Smith2014,Schruba2018,Sternberg2023}.

\citet{Elmegreen1993} suggested that the ISM midplane pressure from dynamical equilibrium models determines the balance of the molecular and atomic phases of the ISM. They find that this pressure term is tightly related to the fraction of molecular gas and thus influences the fraction of ISM in the dense star-forming phase. Follow-up studies examined this and also found a tight relationship between the pressure term and \rmol\ \citep[e.g.][]{Wong2002,Blitz2006,Leroy2008,Sun2020_pde} or the dense gas fraction traced by HCN/CO line ratio\footnote{The dense gas fraction, $f_{\rm dense}$, is traced by the integrated intensity of HCN(1-0) over CO and has been found to correlate with \rmol\ \citep[e.g.][]{Usero2015}} \citep[e.g.][]{Usero2015,Jimenez-Donaire2019}. Similar to these works, we refer to this term as the ``dynamical equilibrium pressure'' (\Pdyn) throughout this paper. The availability of high-quality \hi\ data is the main limiting factor in the above-mentioned analyses, in terms of angular and spectral resolution, sensitivity, and methodology (i.e. fixing \hi\ related parameters). Because of the forthcoming and improved observations that capture the emission of low surface density \hi\, together with the availability of multi-wavelength data, the question arises whether $R_{\rm mol}$ is correlated with any other physical quantities (e.g. star formation rate, surface density, metallicity, radiation field). Assessing these relationships requires compiling both more sensitive \hi\ observations and a suite of multi-wavelength data.

\begin{figure*}[ht!]
    \centering
    \includegraphics[width=1.0\textwidth]{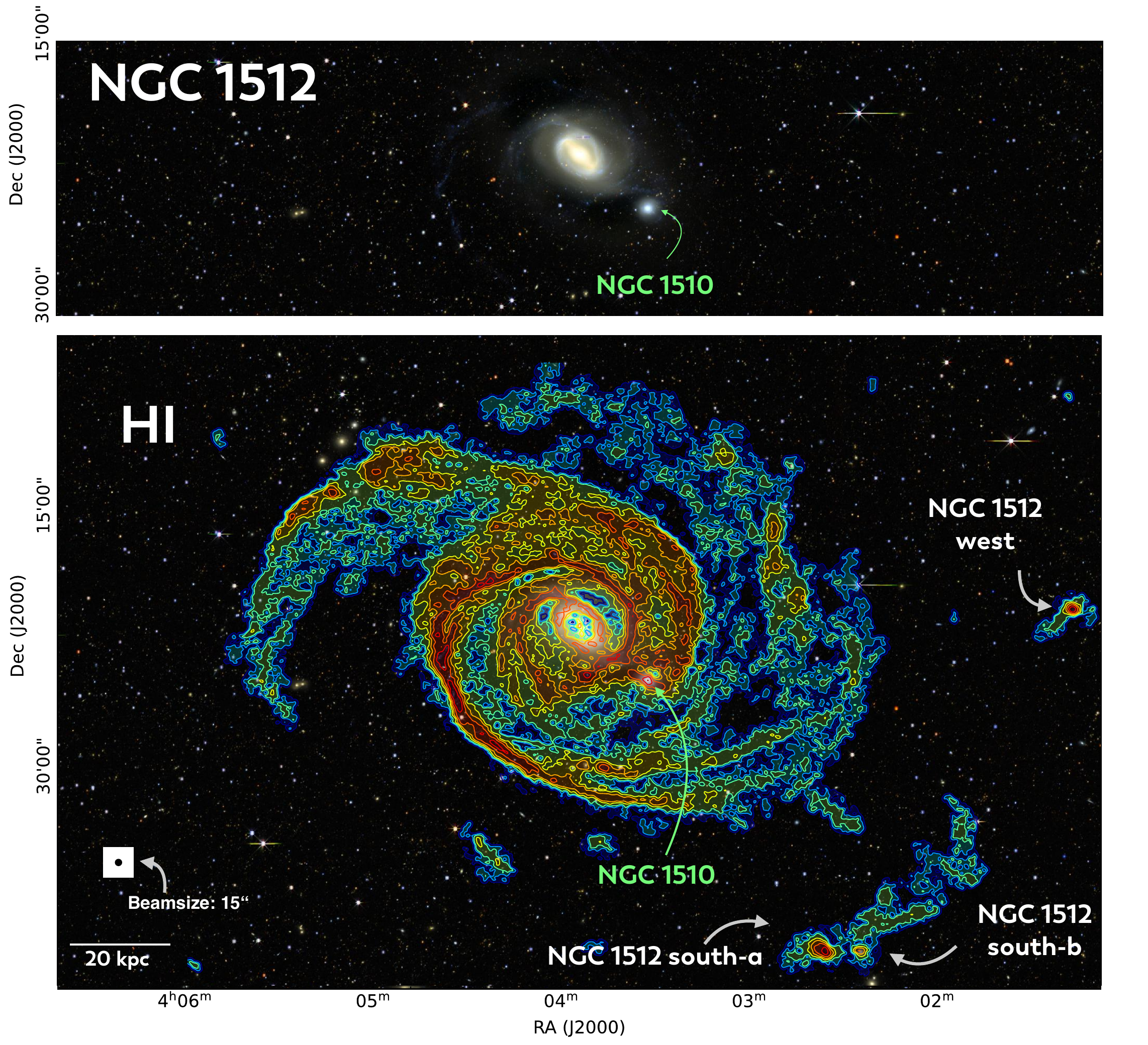}
    \caption{\textit{Top}: Three-color optical image showing star forming spiral galaxy NGC~1512 and the low-mass dwarf companion NGC~1510 in the optical (image credits: DESI Legacy Survey). The scale bar on the lower left corner indicates 20~kpc. \textit{Bottom}: Same as above but overlaid with \hi\ emission contour levels of $\log_{10}$ 0.25, 1, 1.5, 2, 2.25, 2.5, 2.75, 3 K~km~s$^{-1}$ together with the western and southern tidal dwarf galaxies of NGC~1512 (see \autoref{sec:the_sampe} for more information).}
    \label{fig:ngc1512_color}
\end{figure*}

Recently, many of the key required data have been assembled for a large set of galaxies in the context of the PHANGS\footnote{\url{www.phangs.org}} surveys \citep[see, ][]{Leroy2021_pipe,Leroy2021_survey,Emsellem2022,Sun2022} to revisit this analysis. One of the missing pieces has been high sensitivity \hi\ observations. In this work, we fill this gap with new high-quality \hi\ observations of eight nearby galaxies that are also in the PHANGS-ALMA sample, thus supported by available, comprehensive multi-wavelength data. We analyze new \hi\ observations taken with MeerKAT (the SKA precursor facility, \citealt{Jonas2016}) towards eight nearby galaxies and present the first comparison of new MeerKAT and ALMA observations to measure \rmol. As of 2024, MeerKAT is the most sensitive centimeter-wavelength interferometer in the southern hemisphere. The total MeerKAT antenna gain is $\sim$2.8 K/Jy and the system temperature is $\sim$18~K for the L band (856-1712~MHz) which makes it ideally suited to detect low column density \hi\ emission \citep[see e.g.][]{deBlok2016,Elson2023,Boselli2023,deBlok2024,Healy2024,Serra2019,Serra2023,Maccagni_2024,Laudage_2024}.

In this paper, we measure how the molecular gas fraction (\rmol) depends on local conditions in galaxy disks.  We conduct new measurements of \rmol\ for 8 MeerKAT+ALMA data sets (see \autoref{sec:observation}, \autoref{sec:the_sampe} and \autoref{sec:physical_quantities}) and compare these to the total gas surface density (\siggas), the star formation rate surface density (\sigsfr), the stellar mass surface density (\sigstar), the dynamical equilibrium pressure (\Pdyn), and metallicity (Z($r_{\rm gal}$)) (see \autoref{sec:results} and \autoref{sec:discussion}). This pilot study illustrates the power of MeerKAT and ALMA synergies to study key questions of galaxy evolution and reflects the best set of \rmol\ measurements based on the high sensitivity of these datasets. 

\begin{table*}[]
\centering
\caption[Properties of our compiled dataset]{Properties of our compiled dataset for \sample\ (from PHANGS-MeerKAT; this work) and \samplemhongoose\ (from MHONGOOSE, \citealt{deBlok2016}). We provide in columns from left to right the galaxy name, morphological type (Morph.), central right ascension (RA) and declination (Dec.), distance (Dist.), effective radius ($R
_{\rm eff}$), optical radius ($r_{25}$), inclination ($i$), position angle (P.A.), stellar scale length ($l_{\star}$), the total mass of stars ($M_{\star}$), and global star formation rate (SFR).}
\label{tab:sample_properties}
\begin{tabular}{lccccccccccc}
\hline \hline \\
Galaxy & Morph. & RA     & Dec.      & Dist. & $R_{\rm eff}$ & $r_{\rm 25}$ & $i$  & P.A. & $l_{\star} $ & $M_{\star}$  & SFR \\
& & {[$^{\circ}$]} & [$^{\circ}$] & [Mpc] &  \multicolumn{2}{c}{[kpc]} & [$^{\circ}$] & [$^{\circ}$] & [kpc] & [log(M$_{\odot}$)] & [log($M_{\odot}~\text{yr}^{-1}$)] \\ 
        & (1)       & (2)     & (3)      & (4)  & (5) &  (6)  & (7)   & (8) & (9) & (10) & (11)\\ \hline
IC\,1954  & Sb      & 52.522  & -52.074  & 12.8 & 2.4 & 5.6 & 57.1 & 63.4  & 1.5 &  9.7  & -0.4     \\
NGC\,1512 & SB(r)ab & 60.976  & - 43.349 & 18.8 & 4.8 & 23.1 & 42.5 & 261.9 & 6.2 & 10.7 & 0.1  \\
NGC\,1566 & SABb    & 65.002  & - 54.938 & 17.7 & 3.2 & 18.6 & 29.5 & 214.7 & 3.9 & 10.8 & 0.7 \\
NGC\,1672 & Sb      & 71.427  & - 59.247 & 19.4 & 3.4 & 17.4 & 42.6 & 134.3 & 5.8 & 10.7 & 0.9  \\
NGC\,3511 & SABc    & 165.238 & -22.817  & 13.9 & 3.0 & 12.3 & 75.1 & 256.8 & 2.4 & 10.4 & 0.1     \\
NGC\,4535 & SAB(s)c & 188.585 & 8.198    & 15.8 & 6.3 & 18.7 & 44.7 & 179.7 & 3.8 & 10.5 & 0.3  \\
NGC\,5068 & Sc      & 199.728 & -21.039  & 5.2  & 2.0 & 5.7 & 35.7 & 342.4 & 1.3 & 9.4  & -0.6 \\
NGC\,7496 & SBb     & 347.447 & -43.428  & 18.7 & 3.8 & 9.1 & 35.9 & 193.7 & 1.5 & 10.0 & 0.4 \\
\hline
\end{tabular}
\begin{minipage}{2.0\columnwidth}
        \vspace{1mm}
        {\bf Notes:} 
        (1): The morphological classifications are from HyperLEDA \citep{Makarov2014}.
        (2--3): Central positions are taken from \cite{Salo2015}.
        (4): Source distances are taken from \cite{Anand2021}.
        (5): $R_\mathrm{eff}$ contains half of the stellar mass of the galaxy and is taken from \cite{Leroy2021_survey}. 
        (6): $r_{25}$ is based on the radial position of the isophote at 25~mag~arcsec$^{-2}$ and is taken from \cite{Leroy2021_survey}  that are drawn from RC3 \citep[][]{deVaucouleurs1991} via HyperLEDA.
        (7--8): From \cite{Lang2020}, based on PHANGS CO(2–1) kinematics for the inner disk. 
        (9): The profile-based stellar scale length is taken from \cite{Leroy2021_survey}. 
        (10--11): Derived by \citet{Leroy2021_survey}, using \textit{GALEX}~UV and \textit{WISE}~IR photometry, following a similar methodology to \cite{Leroy2019}.
    \end{minipage}
\end{table*}

\begin{table}[h]
\centering
\caption{Properties of our imaged dataset.}
\begin{tabular}{lccccc}
\hline \hline
Galaxy & \multicolumn{2}{c}{Beam} & $\Delta\nu_{\rm chan}$ & rms & P.c. \\
& [$\arcsec$] & [kpc] & [\kms] & [mJy/beam] & \\
& (1) & (2) & (3) & (4) & (5) \\
\hline
IC 1954  & 15.2 & 0.94 & 1.4 & 0.51 & M\\
NGC 1512 & 15 & 1.36 & 5.5 & 0.22 & P-M\\
NGC 1566 & 12.7 & 1.08 & 1.4 & 0.18 & M\\
NGC 1672 & 15.4 & 1.44 & 1.4 & 0.50 & M\\
NGC 3511 & 14.3 & 1.30 & 1.4 & 0.58 & M\\
NGC 4535 & 15 & 1.14 & 5.5 & 0.28 &P-M\\
NGC 5068 & 13.4 & 0.34 &1.4 & 0.17 & M\\
NGC 7496 & 15 & 1.35 & 5.5 & 0.20 & P-M\\
\hline
\end{tabular}
\begin{minipage}{1.0\columnwidth}
        \vspace{1mm}
        {\bf Notes:} 
        (1--2): The size of the beam in angular and linear scales (adopting distances from \autoref{tab:sample_properties}). We used a \texttt{robust}=0.5 for the \texttt{P-M} galaxies, and \texttt{robust}=1.5 for the \texttt{M} galaxies.
        (3): The channel width, i.e. spectral resolution. 
        (4): Root mean square (rms) noise.
        (5): Project codes: 
        \texttt{P-M} = PHANGS-MeerKAT (cycle 0 observation, MKT-20163); \textit{this work}. 
        \texttt{M} = MHONGOOSE (MKT-SCI-20180515-EB-01) \citep{deBlok2016}. To be able to compare these galaxies, we sample all of them on a hexagonal sampling grid with 1.5~kpc sized apertures (driven by the worst linear resolution in our sample, NGC~1672).
    \end{minipage}
\label{tab:observations_details}
\end{table}

\vfill

\section{MeerKAT 21-cm maps to compare with PHANGS-ALMA CO data}
\label{sec:observation}

%
%

\begin{figure*}[ht]
    \centering
    \includegraphics[width=1.0\textwidth]{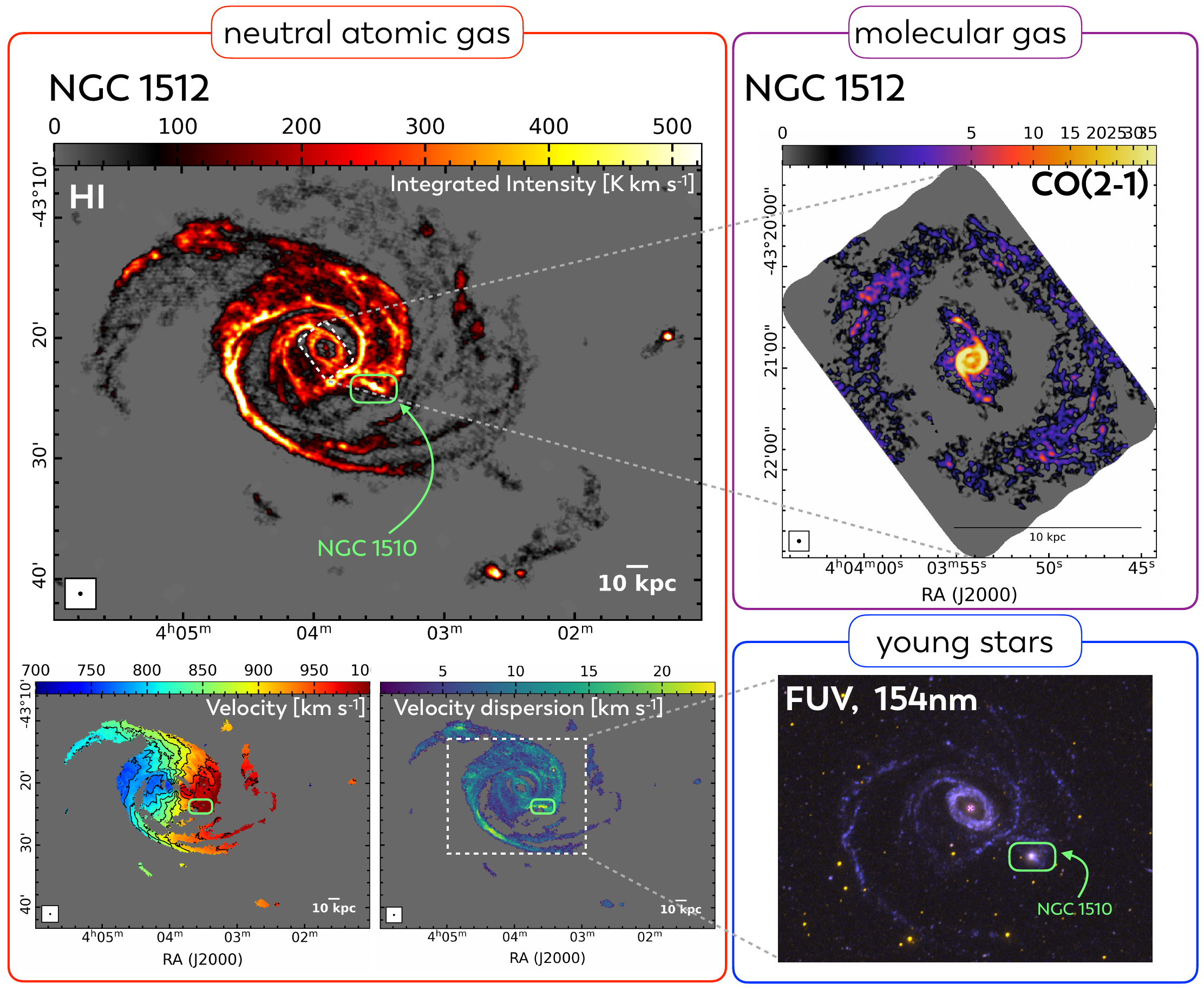}
    \caption{Example of a multi-wavelength view of one of our galaxies within our sample -- NGC\,1512. \textit{Left in red frame:} MeerKAT \hi\ observations showing on top the integrated intensity map in units of K~km~s$^{-1}$ (moment 0), which we use in this work.  Although not used in this paper, the bottom panels show the velocity field (moment 1) and the velocity dispersion (moment 2). These maps are used in \cite{Laudage_2024} to analyze the neutral atomic and molecular gas kinematics in the PHANGS-MeerKAT sample. \textit{Right top in purple frame:} ALMA observations of CO(2-1) from PHANGS-ALMA \citep{Leroy2021_survey} showing an integrated intensity map (see \autoref{sec:2_obs_COcoverage} for CO coverage). \textit{Right bottom in blue frame:} GALEX observation of FUV emission at 154~nm. The green rectangle shows the location of the interacting galaxy NGC~1510.}
    \label{fig:ngc1512}
\end{figure*}

In this work, we analyze new \hi\ observations taken with MeerKAT (the Square Kilometre Array (SKA) precursor facility, \citealt{Jonas2016}) towards 
eight nearby galaxies that are also in the PHANGS-ALMA sample and therefore have multi-wavelength data available. Three of the eight galaxies are from ongoing efforts of the PHANGS collaboration to cover the PHANGS-ALMA sample with MeerKAT (Cycle 0 observations PI: D. Utomo, described in this work). Further, we include five more galaxies from the MeerKAT HI Observations of Nearby Galactic Objects - Observing Southern Emitters (MHONGOOSE, PI: W.J.G. de Blok) sample (\citealt{deBlok2016}).

\subsection{New PHANGS-MeerKAT 21-cm observations of NGC~1512, NGC~4535, and NGC~7496}

The observations were carried out using the South African MeerKAT radio telescope. The data for NGC~1512, NGC~4535, and NGC~7496 were taken between 2021 April 11 and 2021 June 14 for two times 3.5 hours for each galaxy. The observations were taken using the 32k wide mode of the MeerKAT correlator, which provides 26.6~kHz wide channels (corresponding to 5.5 km~s$^{-1}$ at z=0) and has a total bandwidth of 856~MHz. It is thus well suited to resolve the \hi\ line and cover the full range of 21-cm emission from each galaxy. Each of the tracks consists of 10 min of observing time on the primary calibrator J1939-6342 followed by five cycles of two min on a phase calibrator and 3 hours on target.

\subsubsection{Reduction}
The observations of \sample\ were reduced using the CASA-based \texttt{IDIA} calibration pipeline\footnote{\url{https://idia-pipelines.github.io/docs/processMeerKAT}}\citep[see ][]{9560276}. Briefly, the data were put through an initial flagging stage to eliminate regions of known persistent radio frequency interferences (RFI), flag known bad antennas, and identify the strongest RFI outside the known regions. This was followed by an initial round of continuum calibration using the 4K channel data. \footnote{These are the 8 $\times$ frequency-averaged 32K data.} We split out the parallel hand correlations (XX \& YY) and used the primary calibrator to perform the bandpass calibration and flux scale bootstrapping. The secondary calibrator was used for phase calibration. We passed in the broadband model of the primary calibrator to get accurate fluxes across the band. This initial round of continuum calibration was followed by a second round of automated flagging of fainter RFI visible after a single round of calibration. The data were then processed through one final round of continuum calibration.  Note, that with the IDIA pipeline we do not perform frequency-dependent self-calibration. The continuum calibrated target was put through 2 rounds of phase-only self-calibration, using a pyBDSF mask to constrain the deconvolution to avoid artifacts. 
 The first round is a 1 min phase-only solve, and the second round is a 10 min amplitude and phase solve.  The continuum solutions were then transferred on the 32K channel data, that was subsequently used for imaging.

\subsubsection{Imaging}

We ran the calibrated data for the galaxies \sample\ through a standard PHANGS product creation suite (similar to \citealt{Leroy2021_pipe}; PHANGS-ALMA pipeline v3.0\footnote{\url{https://github.com/akleroy/phangs_imaging_scripts}} using \texttt{CASA 5.7.0-134}), building noise cubes, strict masks at each resolution, and combined native+15\arcsec+30\arcsec+60\arcsec broad masks. The \texttt{PHANGS pipeline} essentially uses the \texttt{multiscale} clean (with scales of 0, 30, 100, and 300\arcsec) followed by a more closely restricted single scale clean. We set \texttt{robust} = 0.5 and \texttt{pblimit} = 0.125 during imaging. 

Prior to that, we used an order 1 baseline to subtract the continuum in the \uv-data, excluding the region around the spectral line in the galaxy itself. 1st order approximations are not always ideal and produced in our case some spectral curvature for some of the bright sources that could not be fitted with an order 1 \uv\ continuum subtraction. Rather than do anything more aggressive, we blanked the images in regions where a significant continuum was detected in the line-free channels after cleaning. More specifically, we picked the line-free channels manually and flagged all regions where \sn\ > 5 in the integrated map, allowing either positive or negative artifacts (i.e. we flagged absorption-like features too), then we also masked a couple of clearly problematic regions manually. These regions did not overlap the galaxy in any case. The biggest visible imaging artifacts remaining come from a bright continuum source to the southeast of NGC~7496 which shows some low-level artifacts. However, this is not enough to affect the overall noise.

After cleaning following the above procedure, we circularized the beam, which gives a slightly coarser resolution but also improves the signal-to-noise ratio (S/N). Our final working angular resolution for the images is $12\arcsec-15.0\arcsec$. The rms noise is ${\sim}0.20-0.22$~mJy/beam (or ${\sim}0.7-0.8$~K) in the central part of the primary beam for all targets at this resolution (see \autoref{tab:observations_details}). Although it is in principle possible to make higher resolution images with adjusted weighting in the future, a resolution of ${\sim}15\arcsec$ that corresponds to ${\sim}1.5$~kpc  in the most distant galaxy is sufficient for our analysis.

\subsection{HI data from MHONGOOSE}
The galaxies \samplemhongoose\,  are part of the MHONGOOSE survey \citep[][]{deBlok2016}. The galaxies NGC~1566 and NGC~5068 were observed for ten 5.5 hour tracks for a total of 55 hours, and the remaining galaxies were already observed for 5.5 hours as part of the MHONGOOSE survey \citep{deBlok2024}. Each of the tracks consists of 10 min of observing time on one of the primary calibrators J1939-6342 or J0408-6545. This is then followed by five cycles of two min on a phase calibrator and $\sim 55$ mins on the target galaxy. The \texttt{c856M4k\_n107M} correlator mode was used, allowing for the 32k-narrow-band and 4k-wide-band to be used simultaneously. The narrow-band mode has 32768 channels of 3.265 kHz width each, giving a total bandwidth of 107 MHz. The data were binned by two channels leading to a measurement set with a channel width of 6.53 kHz (1.4 \kms). 

\subsubsection{Reduction}
The data were processed using the publicly available CARACal data reduction pipeline\footnote{\url{https://github.com/caracal-pipeline/caracal}}. The pipeline flags calibrators for RFI, derives and applies the cross-calibration, and flags the target. This is followed by self-calibration using the continuum. After applying the self-calibration solutions, the sky model derived in these steps is subtracted from the measurement set. A second round of uv-plane continuum subtraction using the line-free channels is then used to remove any residual continuum.

\subsubsection{Imaging}
Data cubes at various resolutions and weightings were created using \texttt{wsclean}  \citep[][]{Offringa2014}. Lower-resolution cubes were used to produce clean masks for the higher resolution cubes using the SoFiA-2 source finder  \citep[][]{Westmeier2021}. The data presented here use a robustness parameter of 0.5 and 1.5, resulting in an average beam size of 11\arcsec\ and 30\arcsec, respectively. For our analysis, we use the higher resolution product (see \autoref{tab:observations_details}).
Moment maps were created with SoFiA-2 using the smooth+clip algorithm. We adopted spatial kernels at 0 and 1 times the beam size (i.e. 0 means no smoothing was applied) and spectral kernels of 0, 9, and 25 channels. In all cases, a source finding threshold of 4 sigma was used. A reliability value of 0.8 was used, but we found that in most cases this had little impact as the signal was well separated from the noise.

\begin{figure*}[ht]
    \centering
    \includegraphics[width=0.9\textwidth]{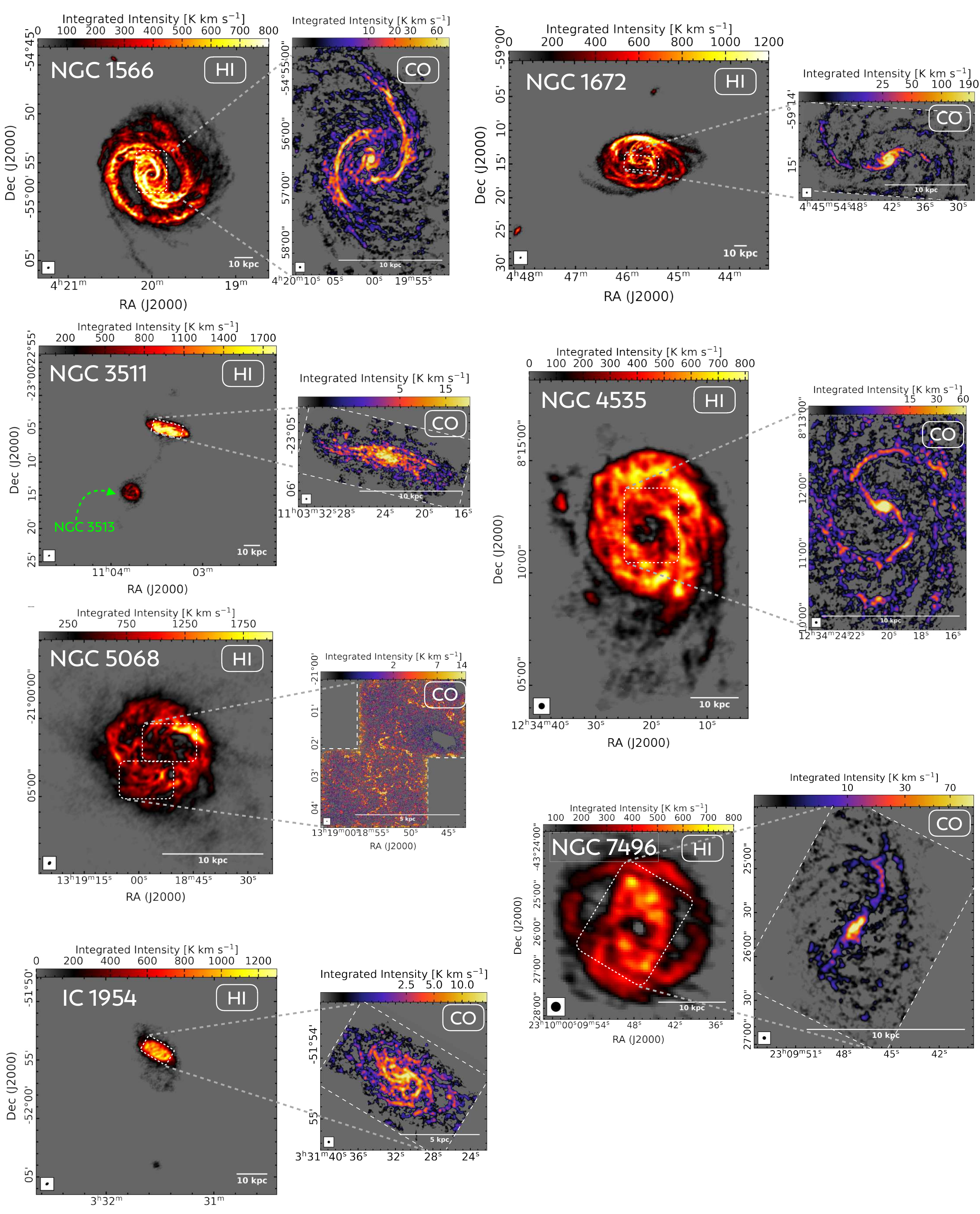}
    \caption{The integrated intensity maps (moment 0) for \hi\ emission (in red to yellow colors) and CO(2-1) (in purple to yellow colors) across the disks of our sample of eight nearby spiral galaxies (excluding NGC~1512, already shown \autoref{fig:ngc1512}). The corresponding beam sizes are visible in the bottom left corner of each map. The white dashed contours on the MeerKAT \hi\ maps represent the field of view of the ALMA CO(2-1) observations, which we show next to each HI map. The CO maps cover ${\sim}70-90\%$ of the total CO emission (see \autoref{sec:2_obs_COcoverage})}
    \label{fig:07_sample}
\end{figure*}

\subsection{PHANGS-ALMA CO data and coverage}
\label{sec:2_obs_COcoverage}
In addition to the MeerKAT 21-cm maps (from PHANGS-MeerKAT and the MHONGOOSE survey) introduced above, we use PHANGS-ALMA CO observations to investigate the molecular gas fraction.The PHANGS-ALMA survey maps the CO J$=2-1$ line emission at ${\sim}1\arcsec$ angular resolution covering the area of active star formation in nearby disk galaxies \citep{Leroy2021_survey}. Although most of the CO emission is expected to lie within the targeted region, weak CO emission can extend into the outer disks of galaxies \citep[e.g.][]{Braine2007,Schruba2011,Koda2022}. As mentioned in Section 4.6.1 of \cite{Leroy2021_survey} the PHANGS CO maps usually cover ${\sim}70-90\%$ of the total CO emission. This result is obtained by using the WISE3 (12$\,\mu$m) observations, which show a surprisingly strong correlation with CO intensity, to make an effective aperture correction (see Section 4.6.1 in \citealt{Leroy2021_survey} for more details).

In this work, we use the emission from the low-$J$ rotational transition of carbon monoxide (CO) as an observational tracer of the molecular gas. The use of CO emission as a \htwo\ tracer comes, however, with a caveat. The \htwo\ abundance is also influenced by the cosmic ray (CR) rate. While the CR rate diminishes at high column densities CRs are less affected by attenuation compared to UV radiation. Moreover, the conditions necessary for the formation of \htwo\ and CO are not the same \citep[see e.g.][]{Gong2018}: in equilibrium, the CO abundance depends primarily on the column density (CO requires $A_{\mathrm{V}}>1$), whereas \htwo\ forms more easily at lower $A_{\mathrm{V}}$ because it is self-shielded. The upper limit on the \htwo\ abundance (in self-shielded gas) depends on density through the balance between formation and CR destruction. It is, however, relevant for this work that CO, and not \htwo, determines the cooling rate in the ISM.


\section{The sample}
\label{sec:distribution_hi}
\label{sec:the_sampe}

In this work we use a compiled dataset that contains new observations from the MeerKAT telescope targeting the galaxies \sample\ from the first results of the PHANGS-MeerKAT survey along with the galaxies \samplemhongoose\ from the MHONGOOSE survey (\citealt{deBlok2016}) Taken together, this forms a sample of eight nearby ($D=5.2-19.4$~Mpc) spiral galaxies (see \autoref{tab:sample_properties}). These eight galaxies are also part of the PHANGS-ALMA survey \citep{Leroy2021_survey} and have archival multi-wavelength observations available. In the following, we describe each galaxy shown in \autoref{fig:ngc1512} and \autoref{fig:07_sample} and give a brief literature overview. 

\begin{itemize}
    \item \textbf{IC~1954}: From the MeerKAT integrated intensity map we see bright \hi\ emission ($>250$~K~km~s$^{-1}$) that extends ${\sim}5$~kpc in radius and is surrounded by fainter emission (${\sim}50$~K~km~s$^{-1}$). The CO map reveals some spiral arm-like structures. The \hi\ phase of this galaxy has not been studied in detail so far. 

    \item \textbf{NGC~1512}: We highlight the barred, double-ring galaxy NGC~1512 in \autoref{fig:ngc1512}. Its inner \hi\ distribution reflects the ring like structuring of the gas, which can also be seen in the CO map. More astonishing, however, are the two gigantic (more than $160$~kpc in length\footnote{We roughly measure the length of a spiral arm as 30\arcmin\ long which translates at NGC~1512's distance of 18.8~Mpc to ${\sim}160$~kpc.}) spiral arms and the prominent disturbances of the \hi\ emission in the southern part of the inner $20$~kpc caused by the interacting low-mass dwarf companion NGC~1510, which is bright in \hi\ emission. The southern arm splits at larger galactic radii into three sub-arms. The northern arm, on the other hand, is clumpier in structure and seems to follow only one direction at larger galactic radii(i.e. no splitting into sub-arms). The distribution and kinematics of the \hi\ gas of this system have also been studied by \cite{Koribalski2009} using the Australia Telescope Compact Array (ATCA). However, our map shows much more detail due to the higher resolution of the MeerKAT observations ($15\arcsec$ compared to a synthesized beam size of $\approx88\arcsec$ from ATCA). In addition to the splitting of the southern arm, our map shows additional clumps of \hi\ emission in the south and east. These were characterized as tidal dwarf galaxies (TDG) by  \cite{Koribalski2009}, who named them simply NGC~1512-south and NGC~1512-west, with both showing clear signs of star formation. In our \hi\ map, however, NGC~1512-south appears as two separate emission clumps. Following the nomenclature we name them NGC~1512-south-a and NGC~1512-south-b. Within the PHANGS collaboration a detailed kinematic study of this \hi\ disk together with the ones from NGC~4535 and NGC~7496 (marked as \texttt{P-M} in \autoref{tab:observations_details}) is underway \citep[][]{Laudage_2024}.

    \item \textbf{NGC~1566}: In \hi\, two remarkably prominent spiral arm-like features are visible with corresponding ``inter-arm'' regions. Both of these arms are winding clockwise, somehow forming a natural extension to the arms seen in CO. NGC~1566, also known as the ``Spanish Dancer'', is an active galaxy classified as Seyfert 1 \citep{Alloin1985,Malkan1998}. \hi\ is detected in absorption against the AGN, with corresponding \hi\ mass about $7\times10^6$ M$_{\odot}$  (see \citealt{Maccagni_2024}). Hence, even though depleted with respect to the CO, there is \hi\ in the central regions tracing the neutral atomic counterpart of what was detected by ALMA in \cite{Combes2014}. Within the MHONGOOSE collaboration a detailed study of this \hi\ disk is under way \citep{Maccagni_2024}.

    \item \textbf{NGC~1672}: From the \hi\ map it is hard to identify clear arm-like features. The \hi\ gas seems to be not uniformly distributed, as we find higher integrated intensities along the northern edge of the \hi\ disk. Also, regions within the field of view of the CO observations show higher integrated \hi\ intensities ($> 300~$K~km~s$^{-1}$) at two regions that are the intersections between a bar and the beginning of two spiral arm-like features -- the bar ends. NGC~1672 is known to have a strong bar that is 2.4\arcmin\ in length, which corresponds to 13.5~kpc at its distance of 19.4~Mpc. These bar ends are only faintly detected in CO. High-resolution optical imaging data from the Hubble Space Telescope (HST) reveal that the bar lane splits into two arms each (i.e. four spiral arms in total) with many HII regions and vigorous star formation along them (e.g. \citealt{Jenkins2011,Barnes2022})

    \item \textbf{NGC~3511}: The \hi\ map shows no clear signs of spiral arm-like features. In CO, two very diffuse, thick, and patchy spiral arms are apparent, however, a clear identification is complicated due to the galaxy's high inclination. The neighboring galaxy NGC~3513 is about 10 arcminutes away from NGC~3511. A filamentary structure with low integrated intensity spanning the two galaxies speculatively seems to connect them. Both galaxies have similar distance estimates of 14.19~Mpc (taken from the MHONGOOSE sample table, \citealt{deBlok2024}).

    \item \textbf{NGC~4535}: The \hi\ map reveals compact emission forming a disk ${\sim}20$ kpc in radius. In the outskirts, two spiral arm-like structures are present, forming (similar to the ones in NGC~1566, but not as prominent) a natural extension to the arms seen in CO.  The observed HI spiral arm-like structures in NGC~4535 resemble those typically influenced by ram pressure, leading to the 'unwinding' of spiral arms in cluster galaxies \citep[see ][]{Bellhouse2021}. Given that NGC~4535 is a member of the Virgo Cluster, it is plausible that similar processes are at play in this galaxy as well. 

    \item \textbf{NGC~5068}: The \hi\ disk of NGC~5068 has a diameter of ${\sim}10$~kpc and shows no clear sign of spiral arm-like structures. NGC~5068 is classified as Sc, is the closest galaxy in our sample with a distance of ${\sim}5.2$~Mpc, and has the lowest total molecular gas mass, the lowest total mass of stars and the lowest star formation rate in our sample (see \autoref{tab:sample_properties}). Apart from the properties mentioned in this table, relatively little is known about this galaxy. Within the MHONGOOSE collaboration, a detailed study of this \hi\ disk is was conducted by \citealt{Healy2024}.

    \item \textbf{NGC~7496}: NGC~7496's appearence in \hi\ is somewhat different compared to the other galaxies. Its \hi\ map reveals an inner compact disk that is surrounded by a symmetric ring-like structure that could be the extensions of the inner two spiral arms. These spiral arm like features are even more evident in the CO map. 
\end{itemize}


\section{Derivation of physical quantities}
\label{sec:physical_quantities}

%
%

Our goal in this work is to measure the H$_2$-to-\hi\ ratio, $R_{\rm mol}$, and to compare it to a set of local environmental factors that, theoretically, have a bearing on the balance of atomic and molecular gas. To identify large-scale conditions associated with total molecular abundance in the ISM, we use a variety of different physical quantities: the atomic gas surface density (\sigatom\,, \autoref{sec:quant_sigatom}), molecular gas surface density (\sigmol\,, \autoref{sec:quant_sigmol}), total gas surface density (\siggas\,,\autoref{sec:quant_sigtot}), star formation rate surface density (\sigsfr\,, \autoref{sec:quant_sigsfr}), stellar mass surface density (\sigstar\,, \autoref{sec:quant_sigstar}), dynamical equilibrium pressure (\Pdyn\,, \autoref{sec:quant_pde}), the ratio between the star formation rate per unit area and dynamical equilibrium pressure  (\sigsfr\,/ \Pdyn) as a proxy for the balance between molecular gas formation versus dissociation (see \autoref{sec:quant_ifuv}), and  ISM metallicity (\zprime, normalized to the solar value, \autoref{sec:quant_metal}).

\begin{figure*}[ht!]
    \centering
    \includegraphics[width=1.0\textwidth]{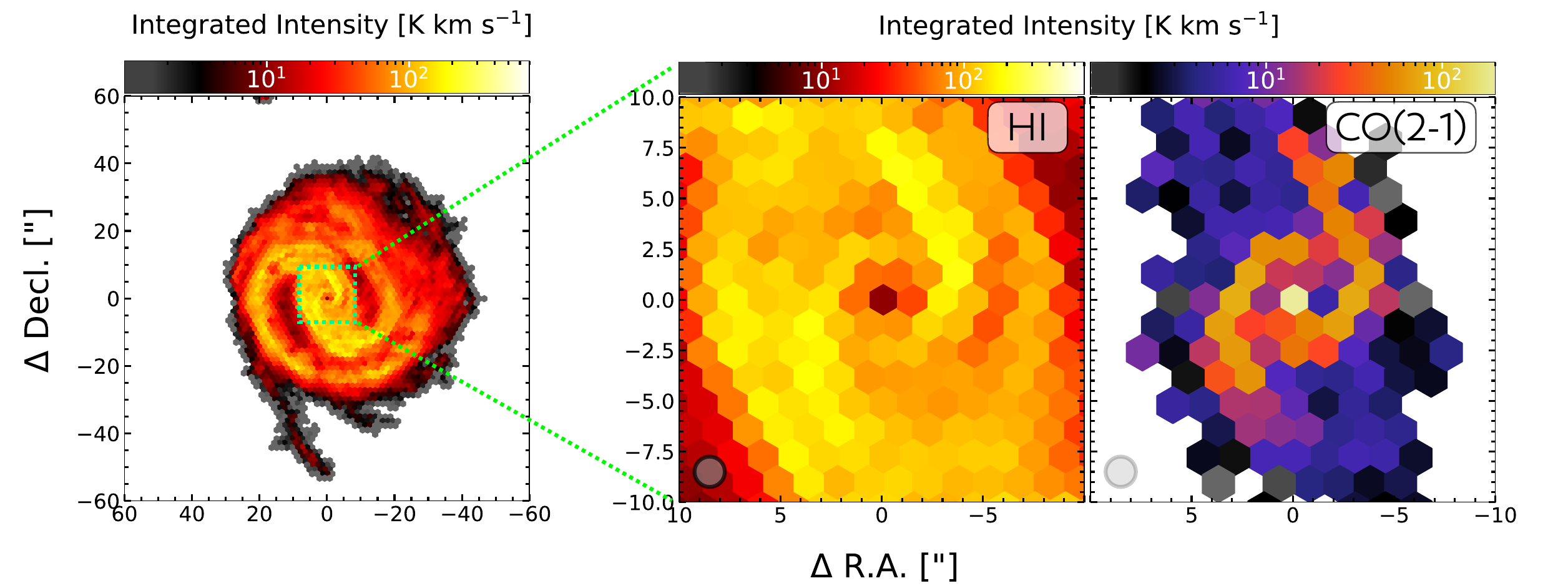}
    \caption[Integrated intensity maps after hexagonal sampling.]{{Integrated intensity maps after hexagonal sampling.} The left plot shows the overall \hi\ disk of NGC~1566. The right two panels show the enlarged versions of the green rectangle at matched field-of-view. We show here hexagonal apertures (1.5 kpc in size) on the \hi\ and CO(2-1) maps where the circle in the lower left corner indicates the (matched) beam size (see \autoref{tab:observations_details}).  }
    \label{fig:sample_matched_resolution}
\end{figure*}

We extract measurements in a set of 1.5~kpc hexagonal apertures\footnote{We chose 1.5~kpc sized hexagonal apertures because 1.5~kpc represents one linear resolution element of our \hi\ observations. 1.5~kpc is also the best common resolution achievable for all galaxies in our sample.} \,that completely tile the sky footprint of each galaxy. This is the same sampling scheme used in \cite{Sun2022} and allows us to leverage a rich set of existing measurements made available by that work (see, for example, \autoref{fig:sample_matched_resolution}). We note that this approach does not require us to convolve the data first to a common beamsize using a Gaussian kernel. We refer interested readers to \cite{Sun2022} for more information. Below we briefly describe how we derive various physical properties.

\subsection{Atomic gas surface density} \label{sec:quant_sigatom}
Assuming optically thin \hi\ emission, we convert \Ihi\ from the MeerKAT observations (see \autoref{sec:observation}) to atomic gas surface density \sigatom\ via,
    \begin{equation}
        \sigatom~[\uSig] = 2.0\times10^{-2}~I_{\rm HI}~[\uIhi]~\cos{(i)}~.
    \label{eq:SigHI}
    \end{equation}
\noindent Here \sigatom\ includes the (extra 35\%) mass of helium and heavier elements. The $\cos{(i)}$ term accounts for galaxy inclination taken from \autoref{tab:sample_properties}. A note of caution is appropriate, as the inclination angle has been seen to vary in the outskirts of galaxies induced by significant warps and tidal effects from infalling galaxies or gas streamers. The main results of this paper should not be impacted by the assumption of a constant inclination angle. In the future, these new maps will yield additional constraints on the inclination and the position angle (for example, \citealt{Laudage_2024} for NGC~1512, NGC~4535, and NGC~7496).

\subsection{Molecular gas mass surface density} \label{sec:quant_sigmol}
We estimate the  molecular gas mass surface  density, \sigmol, from the CO integrated intensity taken from the PHANGS-ALMA survey \citep[][]{Leroy2021_survey} via,
    \begin{equation}
    \sigmol~[\uSig] = \alphaCOline{1}{0}~\Rsub{21}^{-1}~\ICO~[\uIhi]~\cos{(i)}~.
    \label{eq:SigCO}
    \end{equation}
\noindent For the CO line ratio ($\Rsub{21}$) and \CO10-to-\htwo\ conversion factor ($\alphaCOline{1}{0}$) we adopt $\Rsub{21}=0.65$ from \citet{denBrok2021} and \citet{Leroy2022} and use a metallicity-dependent $\alphaCO$ prescription as described in \citet{Accurso2017} and \citet{Sun2020} as $\alphaCOline{1}{0} = 4.35~Z\spsc{\prime-1.6}~\mathrm{M_\odot~(\uIco~pc^2)^{-1}}$, where $Z'$ is the inferred local gas phase O/H abundance normalized to the solar value (see \autoref{sec:quant_metal})\footnote{We note that this Z-based prescription overall matches with the velocity dispersion prescription derived by \citet{Teng2024}, although the former has a ${\sim3\times}$ larger uncertainty. Additionally, the $\alphaCOline{1}{0}$ is likely to be overestimated in the centers of the galaxies with the $Z'$-based prescription. Our central regions are 1.5~kpc in size and most of this effect would most likely be averaged out. To be able to compare with literature trends, we decided to use the $Z'$-based prescription.}. Again, \sigmol\ includes the (extra 35\%) mass of helium and heavier elements. The same $\cos{(i)}$ inclination correction from Equation~\ref{eq:SigCO} also applies here.

\subsection{Total gas surface density and molecular gas fraction} \label{sec:quant_sigtot}\label{sec:quant_frac}
We estimate the total gas surface density, \siggas\ from the sum of \autoref{eq:SigHI} and \autoref{eq:SigCO} as,
\begin{equation}\label{eq:siggas}
     \siggas~[\uSig] = \sigtot.
\end{equation}
We estimate the  molecular gas fraction, \rmol~=~\htwo-to-\hi~, by taking the ratio as,
\begin{equation}\label{eq:rmol}
  R_{\rm{mol}} = \frac{\sigmol~[\uSig]}{\sigatom~[\uSig]}\;.
\end{equation}
For measuring the molecular gas fraction, we include non-detections. For the case of no significant detection (i.e. below a signal-to-noise ratio of 3) in \sigmol\, we replace the values with upper limits (3 times the uncertainty) and include the propagated errors.  We report for only one 1.5~kpc sized aperture in one galaxy (NGC~7496, see \autoref{sec:radial_extent}) a non-detection in \sigatom\ where we detect \sigmol, and thus, we do not include lower limits in our \rmol\ measurements. 

\subsection{Star formation rate surface density} \label{sec:quant_sigsfr}
We estimate the star formation rate (SFR) surface density \sigsfr\ by combining ultraviolet (UV) data from the Galaxy Evolution Explorer (GALEX; \citealt{Martin2005}) and mid-infrared (mid-IR) data from from the Wide-field Infrared Survey Explorer (WISE; \citealt{Wright2010}), following a new SFR calibration \citep[][]{Belfiore2023}. In detail, we make use of the FUV 154\,nm, near-IR 3.4$\,\mu$m (WISE1), and mid-IR 22$\,\mu$m (WISE4) data. These datasets were processed as part of the z0MGS project \citep[][]{Leroy2019}.

 Building upon \cite{Leroy2019}, \cite{Belfiore2023} refined the FUV+WISE4 star formation rate prescription by introducing an additional dependence on the local FUV-to-WISE1 color. This refinement corrects for contamination by IR cirrus in WISE4 bands, leading to better agreement in the SFR estimates from FUV+WISE4 versus those from extinction-corrected H$\alpha$. The refined prescription can be expressed as
 \begin{equation}
 \label{eq:sigsfr}
    \sigsfr~[\uSigSFR] 
    = C_{\rm FUV} L_{\rm FUV} + C_{\rm W4}^{\rm FUV} L_{\rm W4}\;,
 \end{equation}
 where $C_{\rm FUV}$ is the FUV conversion factor established by \cite{Leroy2019}, that is based on SED fitting results of \cite{Salim2018}, using a \cite{Chabrier2003} IMF. $C_{\rm W4}^{\rm FUV}$ is the conversion factor in units of $M_{\odot}$yr$^{-1}$/(erg s$^{-1}$). In our case, we take their recommendation of $C_{\rm W4}^{\rm FUV}$ depending on $Q\equiv$L$_{\rm FUV}$/L$_{\rm W1}$, a broken power law in the form of,
\begin{equation}
    \log{C_{\rm W4}^{\rm FUV}} = 
    \begin{cases}
     a_0 + a_1~\log{Q} & Q < Q_{\rm max} \\
     \log{C_{\rm max}} & Q > Q_{\rm max} 
    \end{cases}\;,
\end{equation}
where $Q_{\rm max}=0.60$, $a_0$ is $\text{log}~C_{\rm max}-a_1~\log{Q_{\rm max}}$, with
$a_1$ = 0.23, $\log{C_{\rm max}}$=0.60 (see also Table 3 and Equation 7 in \citealt{Belfiore2023}). This description is better at mitigating the IR cirrus contamination. 

We show in \autoref{appendix:sfr} a comparison of our SFR measurements with H$\alpha$-based SFR measurements that are corrected for dust extinction based on the Balmer decrement \citep[see][]{Belfiore2023} for the 5 out of our 8 galaxies that are in the PHANGS-MUSE sample \citep{Emsellem2022}. We find that the H$\alpha$-based \sigsfr\ show a similar trend but with smaller scatter and thus might reveal a tighter relationship between \rmol\ and \sigsfr. 

\subsection{Stellar mass surface density} \label{sec:quant_sigstar}
We use the stellar mass surface density \sigstar\ maps computed with the technique utilized for the PHANGS-ALMA survey \citep{Leroy2021_survey}. This \sigstar\ estimate is based on near-infrared emission observations at 3.6 $\mu$m (IRAC1 on Spitzer, from the S$^4$G survey; \citealt{Sheth2010}) or 3.4 $\mu$m (WISE1, from the z0MGS project; \citealt{Leroy2019}). The final stellar mass is then derived from the NIR emission using a mass-to-light ratio, $\MtoLwiseone$, that depends on the local specific star formation rate SFR$ /M_{\star}$,
\begin{align}\label{eq:sigstar_3p6um}
    {\sigstar}[\uSig] &= 350 \left(\frac{\MtoLwiseone}{0.5}\right)~{\Iiracone}~[\uI]~ \cos{(i)}~. 
\end{align}

\subsection{Dynamical equilibrium pressure} \label{sec:quant_pde}

In general, the dynamical equilibrium pressure is the expected average ISM pressure near the galaxy mid-plane needed to balance the weight of the ISM (per unit area) in the galaxy's gravitational potential.
We derive the kpc-scale dynamical equilibrium pressure, \Pdyn, in the same way as in \citet{Sun2020_pde} that is based and built on considerable previous work \citep{Spitzer1942,Elmegreen1989,Wong2002,Blitz2004,Blitz2006,Leroy2008,Ostriker2010,Kim2011,Kim2013}. We calculate \Pdyn\ under the assumptions that (i) the distribution of gas and stars in a galaxy disk can be treated as isothermal fluids in a plane-parallel geometry, (ii) the (single component) gas disk scale height is much smaller than the stellar disk scale height, (iii) gravity due to dark matter can be neglected since it represents only a minor component in the inner disk (the relevant region where we can calculate \Pdyn) of massive galaxies. \Pdyn\ can then be expressed as the sum of the weight of the ISM due to its self-gravity \citep[see e.g.][]{Spitzer1942,Elmegreen1989} and the weight of the ISM due to the stellar potential \citep[see e.g.][]{Spitzer1942,Blitz2004}:
\begin{equation} \label{eq:Pde}
    \Pdyn\ [k_\mathrm{B} \Pdynunit] = \left(\frac{\pi\ G}{2}\Sigma^2_{\rm gas}+\Sigma_{\rm gas}\sqrt{2G\rho_{*}}\sigma_{\rm gas,z}\right)~.
\end{equation}

\noindent Here, \siggas\ is the total gas surface density (see \autoref{sec:quant_sigtot}), and $\rho_{*}$ is the stellar mass volume density near the disk midplane that we can describe as
\begin{equation}
    \rho_{*} = \frac{\sigstar}{4~h_{*}} = \frac{\sigstar}{0.54~r_{*}}.
\end{equation}
\noindent This is under the assumption of an isothermal density profile along the vertical direction and a fixed stellar disk flattening ratio $r_{*}/h_{*}$ = 7.3, where $r_{*}$ is the radial scale length of the stellar disk and $h_{*}$ its scale height. We take $r_{*}$ for each galaxy in our sample from \cite{Leroy2021_survey}. These scale lengths have been derived from the S$^4$G photometric decompositions of 3.6 $\mu$m images (IRAC1 on Spitzer; \citealt{Sheth2010}).
\cite{Sun2020_pde} compared this way of calculating $\rho_{*}$ with a flared disk geometry and found that the latter scenario gives lower \Pdyn\ values. In \autoref{eq:Pde}, $\sigma_{\rm gas,z}$ is the mass-weighted average velocity dispersion of the molecular and atomic gas phases  in the direction perpendicular to the disk plane, 
\begin{equation}
    \sigma_{\rm gas,z} = \frac{\sigmol}{\siggas}\sigma_{\mathrm{mol}}+\left(1-\frac{\sigmol}{\siggas}\right)\sigma_{\mathrm{atom}}.
\end{equation}
To be able to compare our pressure estimates to previous works, we follow \cite{Sun2020_pde} and adopt a fixed $\sigma_{\rm atom}$ of 10 \kms\ (taken from \citealt{Leroy2008}), which holds on 1.5 kpc scales. However, we compare in \autoref{appendix:pdyn} our calculated \Pdyn\ estimates with the ones where we let $\sigma_{\rm atom}$ vary and find overall small differences between these two methods, in agreement with \cite{Sun2020_pde}. According to these authors the assumption of a fixed $\sigma_{\rm atom}$ leads to generally higher \Pdyn. For both $\rho_{*}$ and $\sigma_{\rm atom}$ the resulting deviation lies within 0.2~dex. These are minor differences and therefore should not have too much of an impact, which is why in this work we (i) take a fixed stellar disk flattening ratio, and (ii) fix $\sigma_{\rm atom}$.

\subsection{\texorpdfstring{$P_\mathrm{DE}/\Sigma_\mathrm{SFR}$}~as a proxy for physical conditions driving molecular-to-atomic transition} 
\label{sec:quant_ifuv}

Models of photodissociation and cosmic ray (CR) dissociation argue that H$_2$ formation depends on both gas density and UV / CR radiation field \citep[e.g.,][]{Hollenbach1999, Gong2018}. Within a given cloud with density $n_{\mathrm{
cloud}}$, we expect an increase in molecular abundance with $n_{\mathrm{cloud}}/G_0$ or $n_{\mathrm{cloud}}/\xi_{\mathrm{CR}}$ -- the density in cold clouds over the UV radiation field or cosmic ray rate.
Along this line, \cite{Elmegreen1993} considered cloud populations existing under given external ISM pressure and UV radiation field and concluded that the molecular abundance inside these clouds should depend positively on the external pressure and negatively on the radiation field.

In light of these theoretical considerations, here we use the ratio of \Pdyn/\sigsfr\ to capture the combination of key physical parameters determining the molecular gas fraction.
\Pdyn\ represents an observational proxy of the external pressure considered by \citet{Elmegreen1993}; it is also known to correlate strongly with molecular cloud internal pressure \citep{Sun2020_pde}, which in turn follows scaling relations with cloud surface/volume densities  \citep{Sun2018,Sun2020,Rosolowsky2021}.
\sigsfr\ is expected to scale linearly with $G_0$ and $\xi_{\mathrm{CR}}$ as both UV photons and CR come from young stars and stellar remnants created in recent star formation events.
Taken all these together, we expect $\Pdyn/\sigsfr$ to reflect the balance of molecular gas formation versus dissociation, which could drive variations in the molecular gas fraction, \rmol.

\subsection{ISM Metallicity} \label{sec:quant_metal}
Following \cite{Sun2022}, we use in this work the scaling relations to obtain the general trends of  ISM metallicity variation across our sample of eight galaxies. We assume a global galaxy mass--metallicity relation \citep{Sanchez2019} and a fixed radial metallicity gradient within each galaxy \citep{Sanchez2014}, which gives,
\begin{equation}
     \log_{10}\!Z'(r_{\rm gal}) = \log_{10}\!Z'(\Reff) - 0.1\frac{r_{\rm gal} - \Reff}{\Reff}~,
\end{equation}
\noindent where $Z'(r_{\mathrm{gal}})$ is the normalized abundance at arbitrary $r_{\rm gal}$ and $Z'(\Reff)$ is the local gas-phase abundance at $r_{\mathrm{gal}}=\Reff$ normalized by the solar value [$12+\log\mathrm{(O/H)}=8.69$],
\begin{equation}
\begin{split}\label{eq:z_re}
    \log_{10}\!Z'(\Reff) = 0.04 + 0.01\,\left(\log_{10}\!\frac{\Mstar}{M_\odot} - 11.5\right)\\
    \times\exp\left(-\log_{10}\!\frac{\Mstar}{M_\odot} + 11.5\right)~.
\end{split}
\end{equation}
$\Mstar$ in \autoref{eq:z_re} is the galaxy's global stellar mass under the assumption of a Chabrier IMF \citep{Chabrier2003}. Note that these scaling relations are appropriate for abundance measurements adopting the O3N2 calibration in \citet{Pettini2004}. For more details see the Appendix in \cite{Sun2022}. 
We show in \autoref{appendix:metallicity} a comparison of metallicity measurements, where we include metallicities of individual \ion{H}{II} regions for the galaxies that are in the PHANGS-MUSE sample \citep{Groves2023}.

\subsection{Scaling relations and propagated errors}
\label{sec:linmix_error}

We characterize scaling relations by including measurement errors and upper limits, using the hierarchical Bayesian method described in \cite{Kelly2007}. This approach is available as a python package: \texttt{linmix}\footnote{\url{https://linmix.readthedocs.io/en/latest/index.html}}. It performs a linear regression of $y$ on $x$ while incorporating measurement errors in both variables and accounting for non-detections (upper limits) in~$y$. The regression assumes a linear relationship in the form of
\begin{equation}
    \log_{10}(y) = \beta \times \log_{10}(x) + \alpha~,
\end{equation}
where $\beta$ is the slope, and $\alpha$ is the $y$-intercept. We use this approach for \autoref{sec:scaling_relations}. 

We derive the propagated errors ($\sigma_\mathrm{prop}$) of the ratios between quantities ($I_\mathrm{ratio}$) as: 
\begin{equation}
    \sigma_\mathrm{prop} = \frac{|I_\mathrm{Quantity_1}|}{|I_\mathrm{Quantity_2}|} \sqrt{{\left(\frac{\sigma_\mathrm{Quantity_1}}{I_\mathrm{Quantity_1}} \right)}^2 + {\left(\frac{\sigma_\mathrm{Quantity_2}}{I_\mathrm{Quantity_2}}\right)}^2}~.
\end{equation}
The errors are expressed on a base~10 logarithmic scale as:
\begin{equation}
    \sigma_\mathrm{log} = \frac{1}{\ln(10)} \times \left( \sigma_\mathrm{prop} / I_\mathrm{ratio} \right) \approx 0.434 \times \left( \sigma_\mathrm{prop} / I_\mathrm{ratio} \right)~.
\end{equation}


\section{Results}
\label{sec:results}

%
%
%
\begin{figure*}[ht]
    \centering
    \includegraphics[width=1.0\textwidth]{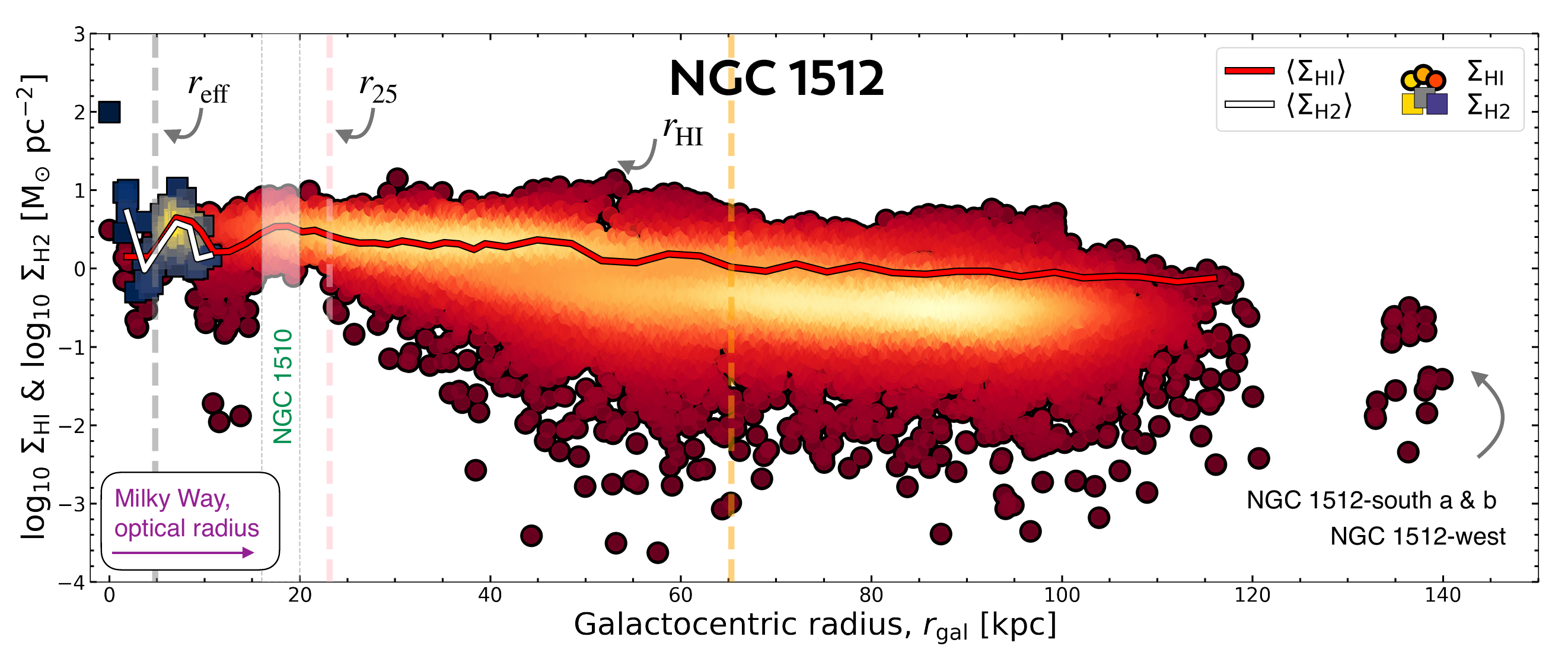}
    \caption{Radial extent of \hi\ and CO in NGC~1512. This galaxy has the largest \hi\ disk in our sample. The maximum galactocentric radius of the CO(2-1) observations is sensitive to the ALMA field-of-view (we recover ${\sim}70-90\%$ of the total CO emission, see \autoref{sec:2_obs_COcoverage}). The scatter points show each individual sight line, i.e. 1.5~kpc aperture. The red-to-yellow color scale shows the point density of \sigatom\ and the dark-blue to yellow shows the point density of \sigmol. The white and red lines show the median within a given 1.5~kpc wide radial bin. We see that \hi\ extends until \rgal\ $\sim120$~kpc (that is $5\times r_{25}$ or 8 times the optical radius of the Milky Way, indicated as a purple arrow). We indicate the approximate \rgal\ range where NGC~1510 sits with vertical dashed lines. The clumps NGC~1512-south-a,  NGC~1512-south-b, and NGC~1512-west are $\sim140$~kpc from the center of NGC~1512. We indicate $R_{\rm eff}$, $r_{25}$ and $R_{\rm HI}$ as grey, pink, and orange vertical dashed lines, respectively (taken from \autoref{tab:sample_properties}).}
    \label{fig:07_radial_extent_ngc1512}
\end{figure*}

\begin{figure*}[ht!]
    \centering
    \includegraphics[width=1.0\textwidth]{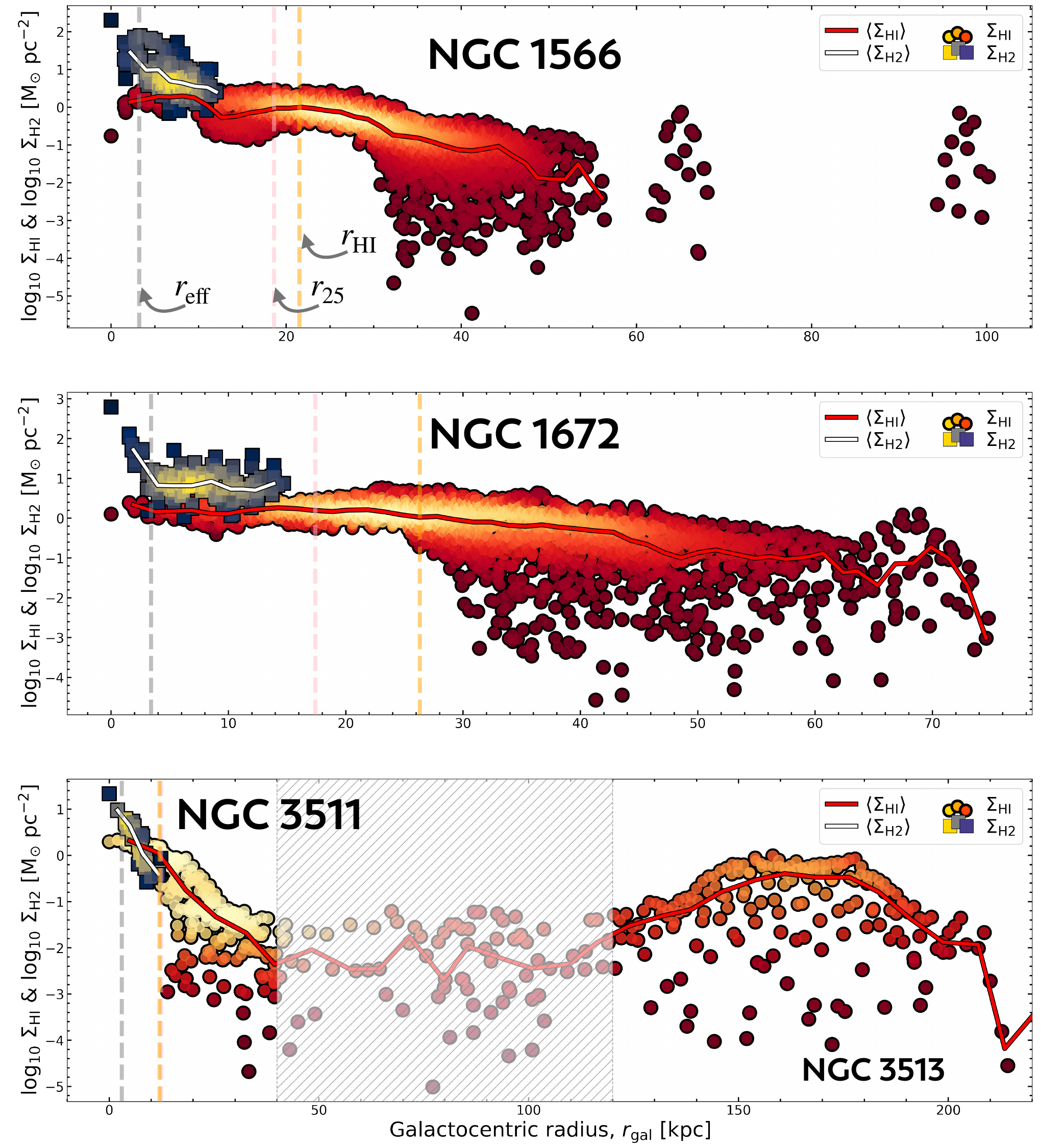}
    \caption{Radial extent of \hi\ and CO in NGC~1566, NGC~1672 and NGC~3511. These galaxies have the largest \hi\ disk extent in our sample, and exhibit special \hi\ characteristics. We indicate $R_{\rm eff}$, $r_{25}$ and $R_{\rm HI}$ as grey, pink, and orange vertical dashed lines, respectively (taken from \autoref{tab:sample_properties}). \textit{Top:} We note here two additional clumps outside of NGC~1566's main \hi\ disk  that have been identified as additional field galaxies \citep[see ][]{deBlok2024}. \textit{Middle:} Toward NGC~1672 we see an increase of \hi\ emission at \rgal\ $\sim70$~kpc, which is the region of additional low-\hi\ emission in the west of NGC~1672's disk (see also \autoref{fig:07_sample}). \textit{Bottom:} The hatched region from \rgal = 40-120~kpc spans the low-density filamentary structure seen in the moment map (see \autoref{fig:07_sample}), which gives the impression of connecting NGC~3511 and NGC~3513.}
    \label{fig:07_radial_extent_big}
\end{figure*}

\begin{figure*}[ht!]
    \centering
    \includegraphics[width=0.95\textwidth]{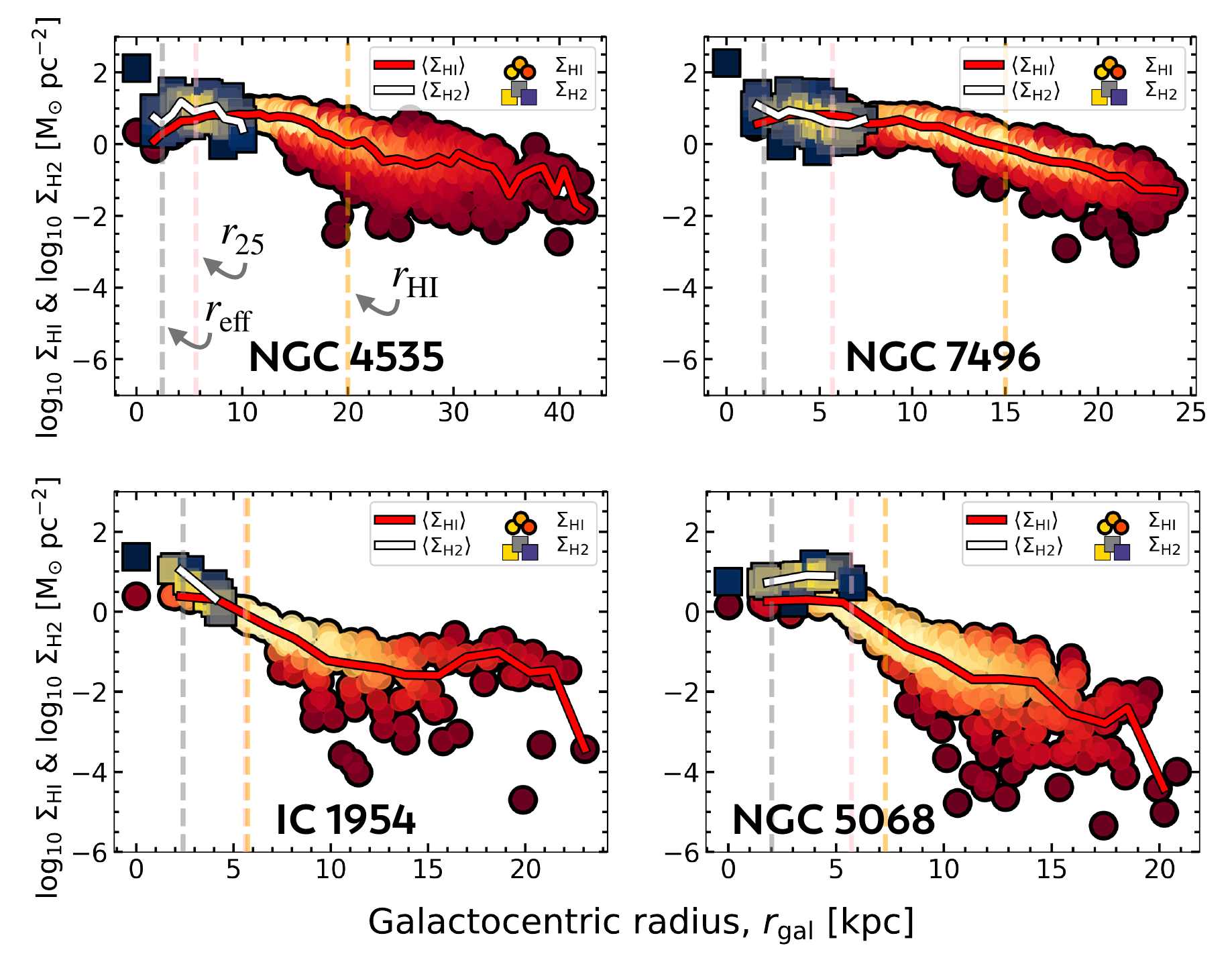}
    \caption{Radial extent of \hi\ and CO in NGC~4535, NGC~7496, IC~1954 and NGC~3511. These galaxies have the smallest \hi\ disk extent in our sample, and do not exhibit any peculiar \hi\ characteristics. We show the corresponding $R_{\rm eff}$, $r_{25}$ and $R_{\rm HI}$ indicated as vertical dashed lines.}
    \label{fig:07_radial_extent_small}
\end{figure*}

\begin{figure*}[ht]
    \centering
    \includegraphics[width=0.95\textwidth]{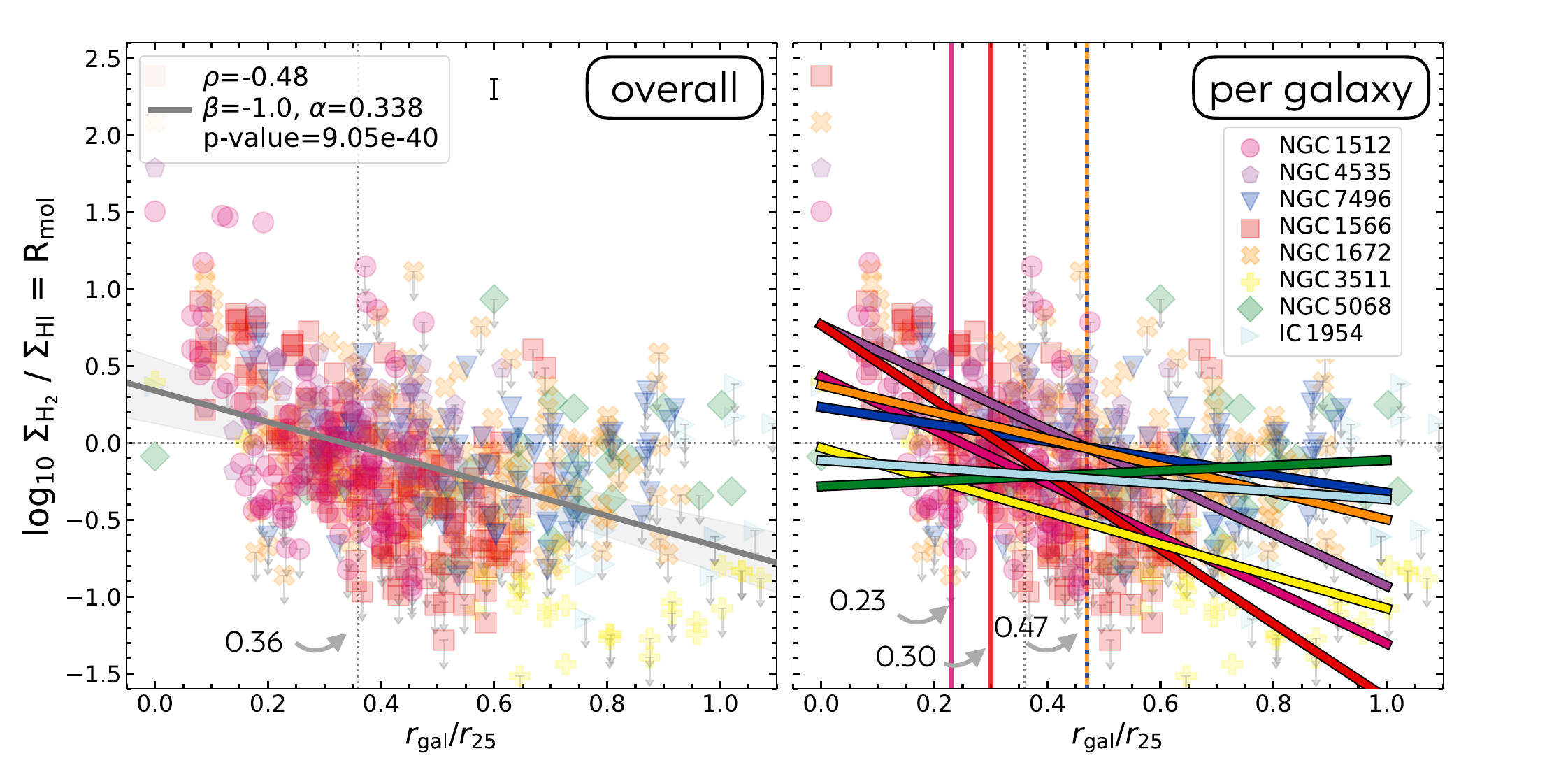}
    \caption[$R_{\rm mol}$ against the normalized radius $r_{\rm gal}/r_{25}$.]{$R_{\rm mol}$ against the normalized radius $r_{\rm gal}/r_{25}$. The region below the dotted horizontal line at $\rmol=1$ can be viewed as dominated by \sigatom, and the region above that line by \sigmol. (\textit{Left}): The grey line shows a fit for all the galaxies and it crosses the $R_{\rm mol}$=1 boundary at $r_{\rm gal}~{\sim}0.4~r_{25}$, which means that the atomic gas on average becomes molecular inside that radius in these eight galaxies. The grey shaded region corresponds to the  3$\sigma$ confidence interval. (\textit{Right}): The colored lines show fits for individual galaxies. We list the corresponding fit parameters in \autoref{tab:07_rmol_r25}. }
    \label{fig:Rmol_vs_r25}
\end{figure*}

\begin{table}
\centering
\caption{Radial extent of our sample.}
\begin{tabular}{lcc|cc|ccc}
\hline \hline
NGC & \multicolumn{2}{c}{$r_{\rm gal,max}$} & \multicolumn{1}{c}{$r_{\rm eff}$} & \multicolumn{1}{c}{$r_{\rm 25}$} &  \multicolumn{3}{c}{$r_{\rm HI}$}\\
or IC & CO & \hi\ &  &  & & & \\
& [kpc] & [kpc] & [kpc] &  [kpc]&  [kpc]& [$r_{\rm eff}$] & [$r_{\rm 25}$]\\
& (1) & (2) & \multicolumn{1}{c}{(3)} & \multicolumn{1}{c}{(4)} & (5) & (6) & (7) \\
\hline
1954  & 5 & 25   &2.4 &  5.6   & 5.7  &${\approx}2$  & ${\approx}1$ \\
 1512 & 10 & 120 &4.8 &  23.1  &  65.3&${\approx} 14$ & ${\approx} 3$\\
 1566 & 12 & 55  &3.2 &  18.6  & 21.5 &${\approx}7$  & ${\approx}1$\\
 1672 & 18 & 75  &3.4 &  17.4  &  26.3 &${\approx} 8$  & ${\approx} 2$\\
 3511 & 13 & 40  &3.0 &  12.3  & 12.0 &${\approx}4$   & ${\approx}1$\\
 4535 & 11 & 45  &6.3 &  18.7  &  20.0 &${\approx} 3$   & ${\approx} 1$\\
 5068 & 6 & 20   &2.0 &  5.7   & 5.8  &${\approx}3$    & ${\approx}1$ \\
 7496 & 8 & 25   &3.8 &  9.1   &  15.0 &${\approx} 4$   & ${\approx}2$ \\
\hline
Aver. & 10 & 51 & 3.6  & 13.8 &  21.5 & $ 6$ &  $2$\\
\hline
\end{tabular}
\begin{minipage}{1.0\columnwidth}
        \vspace{1mm}
        {\bf Notes:} 
        (1--2): The maximum value in galactocentric radius (\rgal) for CO(2-1)\footnote{The maximum galactocentric radius of the CO(2-1) observations are sensitive to the ALMA FoV. We expect, however, that we recover ${\sim}70-90\%$ of the total CO emission (see \autoref{sec:2_obs_COcoverage}).} and \hi, for each galaxy.
        (3-4): $r_{\rm eff}$ and $r_{\rm 25}$ in units of kpc, taken from \autoref{tab:sample_properties}.
        (5): $r_{\rm HI}$ is defined as the \sigatom\ = 1~\uSig\ isophote.
        (6): The value of $r_{\rm HI}$ in units of $r_{\rm eff}$. For example, for NGC~1512 $r_{\rm HI} =  65.3~{\rm kpc} \approx  14 r_{\rm eff}$. 
        (7): $r_{\rm HI}$ in units of $r_{\rm 25}$. For example, for NGC~1512 $r_{\rm HI} \approx 3 r_{\rm 25}$. 
    \end{minipage}
\label{tab:07_r25}
\end{table}

\begin{figure*}[ht!]
    \centering
    \includegraphics[width=0.95\textwidth]{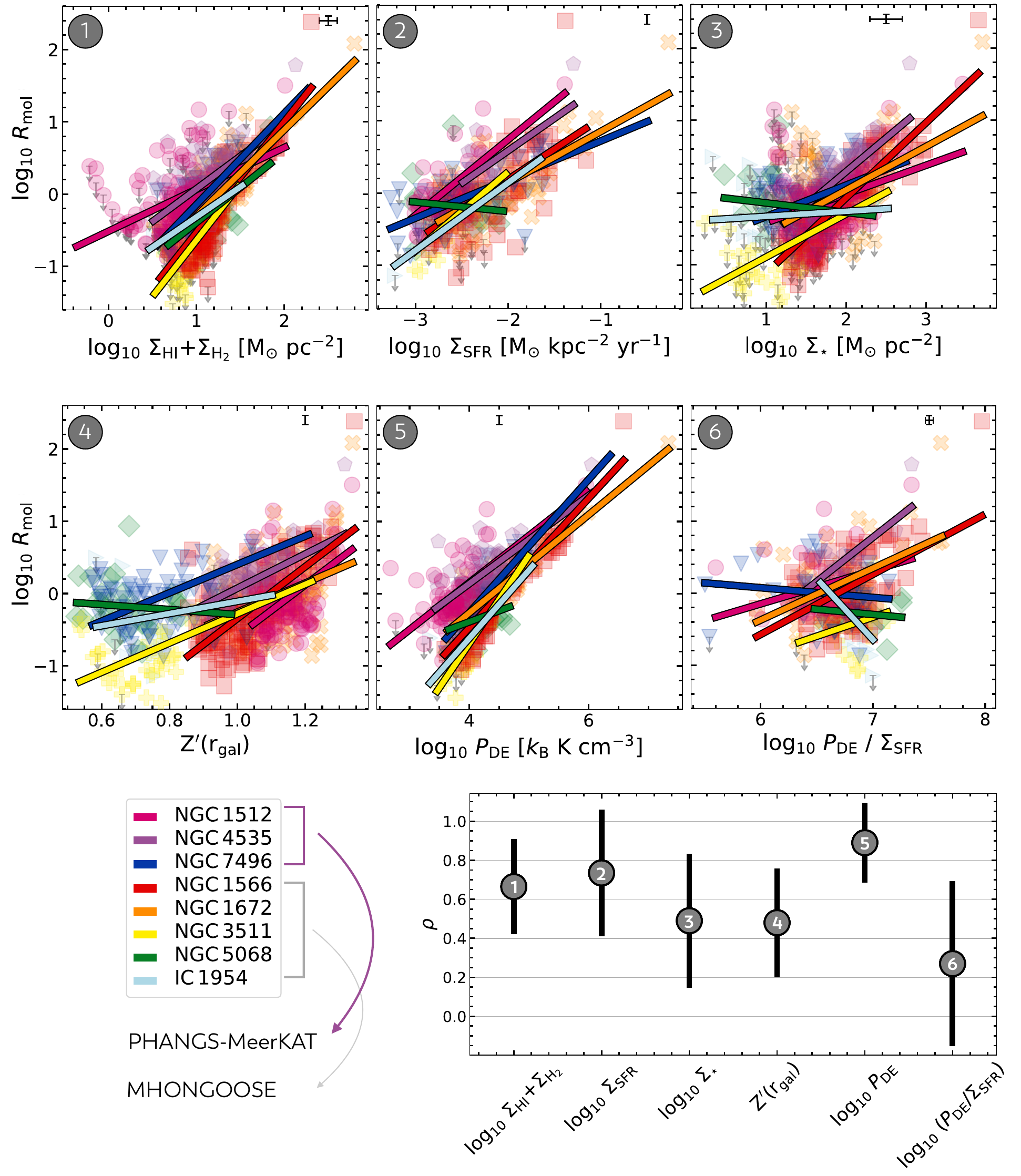}
    \caption{$R_{\rm mol}$ scaling relations. The numbered panels show linear regression fits between log$_{10}$~\sigmol\ /\sigatom\ against six different physical quantities that we derived in \autoref{sec:physical_quantities}: (1) \siggas, (2) \sigsfr, (3) \sigstar, (4) \zprime, (5) \Pdyn, and (6) \Pdyn / \sigsfr. Each color represents one galaxy. The lower right panel represents the range of correlation coefficient $\rho$ and the mean absolute $\rho$ for all galaxies for each individual fitted quantity shown in panels (1) - (6) using a hierarchical Bayesian method (\texttt{linmix}, see \autoref{sec:linmix_error}). All of the properties of the presented linear regression fits are listed in \autoref{tab:correlations}.}
    \label{fig:07_scalingrelations}
\end{figure*}

\noindent In this section, we present the results from the MeerKAT and ALMA observations of our sample of eight nearby galaxies (from the PHANGS and MHONGOOSE surveys). \autoref{fig:ngc1512} and \autoref{fig:07_sample} show the \hi\ and CO moment maps of our sample. We describe the radial extent of \hi\ and CO emission (see \autoref{fig:07_radial_extent_ngc1512}, \autoref{fig:07_radial_extent_big}, \autoref{fig:07_radial_extent_small} and \autoref{fig:Rmol_vs_r25}) in \autoref{sec:radial_extent} and \autoref{sec:rmol_vs_rgal}. Using ancillary data (described in \autoref{sec:physical_quantities}) we analyze \rmol\ scaling relations (see \autoref{fig:07_scalingrelations}) in \autoref{sec:scaling_relations}.

\subsection{Radial extent of the surface density of the atomic and molecular gas}
\label{sec:radial_extent}

In \autoref{fig:07_radial_extent_ngc1512}, \autoref{fig:07_radial_extent_big} and \autoref{fig:07_radial_extent_small} we show the radial extent of \sigatom\ and \sigmol\ together with their point densities, with galaxies grouped by their \hi\ disk extent and special \hi\ characteristics. To derive radial trends we bin \sigatom\ and \sigmol\ in galactocentric bins of 1.5 kpc width (the common beam size, i.e. the size of our hexagonal apertures). The white and red lines thus represent the median \sigmol\ and \sigatom\, respectively, within a given bin. 

Carefully inspecting each individual galaxy map reveals additional clumps, dips, or peaks at larger \rgal, and/or companion galaxies. In \autoref{fig:07_radial_extent_ngc1512} showing NGC~1512, \hi\ extends out to \rgal\ $\sim120$~kpc (that is $8\times$ the radius of the Milky Way, indicated as a purple arrow). NGC~1512's interacting companion (NGC~1510) is indicated by a grey shaded region at \rgal\ $=16--20$~kpc. The clumps NGC~1512-south-a,  NGC~1512-south-b, and NGC~1512-west are $\sim140$~kpc from the center of NGC~1512 and are most likely tidal dwarf galaxies. Additionally, we see an increase in both \sigatom\ and \sigmol\ at \rgal\ ${\sim}8~$kpc, which is the location of the ring-like feature that can be seen in \autoref{fig:ngc1512}. 

In \autoref{fig:07_radial_extent_big} we show the radial extent of NGC~1566, NGC~1672, and NGC~3511. In NGC~1566 \hi\ extends until \rgal\ $\sim55$~kpc with two additional clumps at \rgal\ $\sim65$~kpc and \rgal\ $\sim97$~kpc,  that have been identified as additional field galaxies, LEDA 414611 and LSBG F157-052 \citep[see ][]{deBlok2024}. For NGC~1672 \hi\ extends until \rgal\ ${\sim}65$~kpc. The \hi\ extent in NGC~3511 is dominated by other features besides the steep radial decline in \sigatom. The hatched region indicates the filamentary structure with low integrated intensity, followed by the companion galaxy NGC~3513. 

In \autoref{fig:07_radial_extent_small} we show the galaxies in our sample that have the least extended \hi\ disks and/or do not exhibit special characteristics such as clumps or connecting branches. NGC~4535 reaches $r_{\rm gal}{\sim}40~$kpc. IC~1954, NGC~5068 and NGC~7496 show a maximal \hi\ extent of ${\sim}25~$kpc. We do not detect any \hi\ emission in the inner 1.5~kpc for NGC~7496. This can also be seen in the moment 0 map in \autoref{fig:07_sample}. IC~1954, NGC~5068 and NGC~7496 are those with the lowest total stellar mass ($M_{\star}$) in our sample (see \autoref{tab:sample_properties}).

Inspecting environmental differences reveals that almost all of our galaxies show higher \sigmol\ than \sigatom\ in their central 1.5~kpc (except NGC~7496 as aforementioned). We find that \sigatom\ in these central regions does not surpass $\sim8~\uSig$ \citep[similar to what has been found in][]{Bigiel2008}. NGC~1566 and NGC~1672 are the galaxies that stand out the most in our sample, each having \sigmol\ ${\sim}3$ orders of magnitude higher than \sigatom\ in their central regions.

For our sample, we find a range of maximum radial extent of \sigatom\ of $r_\mathrm{gal,max}$ ${\approx}20-120~$kpc, and for \sigmol\ we find $r_\mathrm{gal,max}$ ${\approx}5-18~$kpc. The measurements of $r_\mathrm{gal,max}$ are based on the detection threshold of at least 3$\sigma$. The \hi\ emission is on average 5 times more extended than the detected CO emission (see also \autoref{tab:07_r25}). We remind the reader that the maximum galactocentric radius of the CO(2-1) observations is sensitive to the ALMA FoV. We expect, however, that we cover ${\sim}70-90\%$ of the total CO emission (see \autoref{sec:2_obs_COcoverage}). 
It is common to use the \sigatom\ = 1~\uSig\ isophote definition for the maximal radial extend of \sigatom, which we indicate as $r_{\mathrm{HI}}$ in \autoref{tab:07_r25} and find $r_{\mathrm{HI}}$ ${\approx}6-65~$kpc. We find that on average $r_{\mathrm{HI}}$ is approximately twice the radial position of the B-band isophote at 25 mag arcsec$^{-2}$ (i.e. $r_{\mathrm{HI}}\approx2r_{25}$) and six times the radius that contains half of the stellar mass of a galaxy (i.e.  $r_{\mathrm{HI}}\approx6r_{\mathrm{eff}}$). Notably, NGC~1512 has a much larger \hi\ extent than this average trend, with  $r_{\mathrm{HI}} \approx3r_{25}$ or  $r_{\mathrm{HI}}\approx14r_{\mathrm{eff}}$.

\subsection{The transition from atomic to molecular gas dominated regimes}\label{sec:rmol_vs_rgal}
We aim to measure at what radius in our galaxies the overall molecular gas surface density is equal to that of the atomic gas. This represents the boundary between the regions where either \hi\ or \htwo\ dominates. To compare galaxies with different stellar radial extents, it is common to normalize the galactocentric radius (\rgal) with $r_{25}$. In \autoref{fig:Rmol_vs_r25} we show \rmol\ versus $r_{\rm gal}$/$r_{25}$. The rainbow-colored symbols show individual 1.5~kpc apertures within each galaxy.  We note that all the galaxies within our sample contribute with a similar amount of individual aperture measurements (median number of sightlines = 115). In the left panel, the gray linear regression fit shows the correlation of \rmol\ and $r_{\rm gal}$/$r_{25}$  in the form of $\log_{10}(y) = \beta \times (x) + \alpha~$ across all galaxies, which we find to be 
\begin{equation}\label{eq:rmol_r25}
   \log_{10} \rmol =  -1.0\pm0.07~({r_{\rm gal}}/{r_{25}})+0.34\pm0.04. 
\end{equation}
The fit crosses the \rmol\ = 1 boundary (in log scale indicated as a horizontal line at 0) at $r_{\rm gal} \sim 0.4~r_{25}$. However, looking at individual galaxies in the right panel of \autoref{fig:Rmol_vs_r25}, such as NGC~1512 or NGC~1566, reveals that for these objects \sigatom\ already dominates at a smaller $r_{\rm gal}$/$r_{25}$, whereas in other galaxies \rmol\ crosses the $\rmol=1$ boundary at larger $r_{\rm gal}/r_{25}$ or always falls below the $\rmol = 1$ line (e.g.~NGC~5068) (see \autoref{tab:07_rmol_r25} for the details about the individual fits). We will discuss this further in \autoref{sec:discussion}.

\begin{table}
\centering
\caption[Radial extent of our sample]{Radii at which galaxies cross the \rmol\ =~1 boundary (i.e., where \sigatom$\approx$\sigmol), ordered by $r_{\rm gal} / r_{\rm 25}$.}
\begin{tabular}{lcccc}
\hline \hline
Galaxy  & $r_{\rm gal}/r_{\rm 25}$ & $\beta$ & $\alpha$ & Scatter   \\
& (1) & (2) & (3) & (4)  \\
\hline
NGC 1512 & 0.23 & -1.70  $\pm$ 0.36& 0.43   $\pm$ 0.11& 0.16\\
NGC 1566 & 0.30 & -2.40  $\pm$ 0.21& 0.77   $\pm$ 0.09& 0.15\\
NGC 1672 & 0.47 & -0.87  $\pm$ 0.16& 0.37   $\pm$ 0.09& 0.17\\
NGC 4535 & 0.47 & -1.70  $\pm$ 0.21& 0.77   $\pm$ 0.08& 0.07 \\
NGC 7496 & 0.47 & -0.56  $\pm$ 0.20& 0.23   $\pm$ 0.12& 0.11\\
NGC 3511 & -    &  -1.00 $\pm$ 0.17& -0.03  $\pm$ 0.12& 0.11\\
NGC 5068 & -    &  0.17  $\pm$ 0.26& -0.28  $\pm$ 0.17& 0.11\\
IC 1954  & -    & -0.25  $\pm$ 0.26& -0.11  $\pm$ 0.20& 0.17\\
\hline
all  & 0.36 & -1.00 $\pm$ 0.07 & 0.34 $\pm$ 0.04 & 0.18 \\ 
\hline
\end{tabular}
\begin{minipage}{1.0\columnwidth}
        \vspace{1mm}
        {\bf Notes:} 
        (1): The normalized radius $r_{\rm gal}/r_{\rm 25}$ at which $\rmol=1$. For example, \sigatom\ starts to dominate at $r_{\rm gal} > 0.23~r_{\rm 25}$ for NGC~1512. (2): The slope $\beta$ of the fit, and (3): The intercept $\alpha$ of the fit. (5): The regression intrinsic scatter. 
    \end{minipage}
\label{tab:07_rmol_r25}
\end{table}

\subsection{\texorpdfstring{$R_{\mathrm{mol}}$} scaling relations}
\label{sec:scaling_relations}

In this work, we investigate how the molecular gas fraction, \rmol, depends on the local conditions in galaxy disks. For this purpose, we include upper limits when calculating \rmol\ (see \autoref{sec:quant_frac}).
In \autoref{fig:07_scalingrelations}, we show for all galaxies the scaling relations of \rmol\ against six physical quantities: the total gas surface density (\siggas), star formation rate surface density (\sigsfr), stellar mass surface density (\sigstar), dynamical equilibrium pressure (\Pdyn), a proxy for ($G_0$/$n$)$^{-1}$ (\Pdyn/\sigsfr), and metallicity (\zprime($r_{\rm gal}$) ). The fits show a linear regression in the form of $\log_{10}(y) = \beta \times \log_{10}(x) + \alpha$, except for \zprime($r_{\rm gal}$) as this quantity is already in logarithmic form. The rainbow-colored markers show the individual aperture measurements in all galaxies.  We note that all the galaxies within our sample contribute with a similar amount of individual aperture measurements (median number of sightlines = 115). Our sample spans almost 4 orders of magnitude in \rmol. We see a range in $\log_{10}(x)$ of ${\sim}1$~dex in panel \zprime and \Pdyn/\sigsfr, ${\sim}3$~dex in \siggas\ and \sigsfr, and ${\sim}4$~dex in \sigstar\ and \Pdyn. 

We find for most of the galaxies moderate\footnote{Here we refer to a ``moderate'' correlation if the correlation coefficient $\rho$ lies in the range of $0.5 - 0.7$ and a ``tight'' correlation if $\rho$ > 0.7.} sub-linear (i.e. $\beta < 1$) correlations between $\mathrm{log}_{10}$~\rmol\ and $\mathrm{log}_{10}$($x$) (see \autoref{tab:correlations} for an overview of correlation coefficient, slope, intercept, and their uncertainties, p-values, covariances and intrinsic scatters). We highlight a few notable exceptions to this general trend.
For correlations between $\mathrm{log}_{10}$~\rmol\ and (i) $\mathrm{log}_{10}$~\siggas\ the galaxies NGCs~7496, 1566, and 1672 show a linear (i.e. $\beta > 1$) correlation, and NGC~1512 is the only galaxy that reveals a weak correlation with $\rho=0.12$, (ii) $\mathrm{log}_{10}$~\sigsfr\ we find for NGC 1954 an exceptionally strong correlation ($\rho=0.97$), and for NGC~5068 a negative correlation ($\rho=-0.14$), (iii) $\mathrm{log}_{10}$~\sigstar\ the galaxy NGC~1566 shows a tight correlation ($\rho=0.85$) with a perfectly linear slope ($\beta=1$), (iv) $\mathrm{log}_{10}$~\zprime\ ($r_{\rm gal}$) all galaxies except NGC~5068 show a linear correlation, (v) \Pdyn\ all galaxies except NGC~5068 show a tight correlation, (vi) \Pdyn/\sigsfr\ we find weak correlations with the largest intrinsic scatter. 

The bottom right panel in \autoref{fig:07_scalingrelations} shows the ranges of $\rho$ (black line) and the median $\rho$ of all galaxies (grey circles), which we also indicate in \autoref{tab:correlations} together with the covariance of the x and y-axis and fit parameters. We find the highest median $\rho$ value for 
(i) \Pdyn\ with $\left<\rho\right>=0.89$ together with a high covariance and the smallest intrinsic scatter, followed by  
(ii) \sigsfr\ with $\left<\rho\right>=0.72$, and
(iii) \siggas\ with $\left<\rho\right>=0.67$ together with the smallest intrinsic scatter but highest covariance compared to \sigsfr\ and \Pdyn. We find the highest covariance for \rmol\ and \sigsfr/\Pdyn\ (Cov($\log_{10}(x)$,$\log_{10}(y)$)~=${\sim}$20), followed by \rmol\ and \Pdyn\ (Cov($\log_{10}(x)$,$\log_{10}(y)$)~=${\sim}$10; see \autoref{tab:correlations}). Looking at the covariances of the individual galaxies between \rmol\ and \Pdyn\ or \sigsfr/\Pdyn, it becomes apparent that NGC~1672 shows the highest covariance values (namely Cov($\log_{10}(x)$,$\log_{10}(y)$)~=~13) for \rmol\ and \Pdyn, and NGC~5068 (Cov($\log_{10}(x)$,$\log_{10}(y)$)~=~25) for \rmol\ and \sigsfr/\Pdyn. 

\begin{table*}[ht!]
\centering
\caption{$R_{\rm mol}$ ($\log_{10}$(y)) scaling relations and correlation coefficients ($\rho$) for the discussed physical quantities \siggas, \sigsfr, \sigstar, \zprime($r_{\rm gal}$), \Pdyn, and \sigsfr/\Pdyn\ ($\log_{10}$(x)) for each galaxy (see \autoref{fig:07_scalingrelations}).}
\label{tab:correlations}
\begin{tabular}{lcccccccccc}
\hline\hline
$\log_{10}$(x) & NGC & $\rho$ & $\left<\rho\right>$ & $\beta$ & $\alpha$  & p-value & Covariance &  Scatter \\
& or IC & correlation coeff. &  & slope & intercept & &  & \textit{intrinsic} \\
&  (1)     & (2)      &  (3)  & (4) & & (5) & (6) & (7) \\
\hline
        & 1512 & 0.12 $\pm$ 0.19                   &      &0.16 $\pm$ 0.14 & -0.17 $\pm$ 0.11 & 0.43                   & -0.01                  & 0.23 \\
        & 4535 & 0.61 $\pm$ 6.50 $\times 10^{-10}$ &      &0.91 $\pm$ 0.12 & -0.83 $\pm$ 0.14 & 2.89 $\times 10^{-11}$ & -8.60 $\times 10^{-3}$ & 0.08 \\
        & 7496 & 0.85 $\pm$ 3.60 $\times 10^{-16}$ &      &1.80 $\pm$ 0.15 & -1.90 $\pm$ 0.16 & 2.93 $\times 10^{-19}$ & -0.59                  & 0.09 \\
\siggas & 1566 & 0.85 $\pm$ 1.60 $\times 10^{-38}$ & 0.67 &1.50 $\pm$ 0.08 & -2.10 $\pm$ 0.10 & 9.55 $\times 10^{-44}$ & -0.03                  & 0.08 \\
        & 1672 & 0.82 $\pm$ 2.90 $\times 10^{-29}$ &      &1.20 $\pm$ 0.07 & -1.50 $\pm$ 0.09 & 7.02 $\times 10^{-34}$ & 0.37                   & 0.07 \\
        & 3511 & 0.72 $\pm$ 2.20 $\times 10^{-6 }$ &      &1.60 $\pm$ 0.24 & -2.20 $\pm$ 0.23 & 1.41 $\times 10^{-7 }$ & 0.69                   & 0.10 \\
        & 5068 & 0.56 $\pm$ 7.00 $\times 10^{-3 }$ &      &0.97 $\pm$ 0.33 & -1.40 $\pm$ 0.41 & 8.69 $\times 10^{-4 }$ & 0.90                   & 0.08 \\
        & 1954 & 0.56 $\pm$ 0.01                   &      &0.83 $\pm$ 0.27 & -1.10 $\pm$ 0.28 & 7.13 $\times 10^{-3 }$ & 0.75                   & 0.12 \\
 \hline 
        & 1512 &  0.63 $\pm$ 1.70 $\times 10^{-4}$  &      & 0.92  $\pm$ 0.19 & 2.50  $\pm$ 0.51 & 5.70 $\times 10^{-5}$   & 4.20 & 0.14 \\
        & 4535 &  0.63 $\pm$ 2.90 $\times 10^{-4}$  &      & 0.89  $\pm$ 0.28 & 2.40  $\pm$ 0.63 & 1.90 $\times 10^{-3}$   & 4.90 & 0.12 \\
        & 7496 &  0.74 $\pm$ 1.50 $\times 10^{-9 }$ &      & 0.87  $\pm$ 0.12 & 2.10  $\pm$ 0.29 & 4.98 $\times 10^{-10}$  & 5.00 & 0.20 \\
\sigsfr & 1566 &  0.72 $\pm$ 3.70 $\times 10^{-17}$ & 0.72 & 0.84  $\pm$ 0.08 & 1.90  $\pm$ 0.18 & 1.51 $\times 10^{-17}$  & 5.10 & 0.14 \\
        & 1672 &  0.72 $\pm$ 1.00 $\times 10^{-13}$ &      & 0.71  $\pm$ 0.08 & 1.60  $\pm$ 0.17 & 4.58 $\times 10^{-14}$  & 2.50 & 0.10 \\
        & 3511 &  0.76 $\pm$ 2.50 $\times 10^{-6 }$ &      & 1.10  $\pm$ 0.22 & 2.50  $\pm$ 0.58 & 1.14 $\times 10^{-5 }$  & 2.90 & 0.06 \\
        & 5068 & -0.14 $\pm$ 0.11                   &      & -0.13 $\pm$ 0.24 & -0.51 $\pm$ 0.62 & 0.59                    & 2.20 & 0.12 \\
        & 1954 &  0.97 $\pm$ 1.10 $\times 10^{-11}$ &      & 0.94  $\pm$ 0.07 & 2.00  $\pm$ 0.17 & 8.89 $\times 10^{-13}$  & 2.30 & 0.01 \\
 \hline 
          & 1512 & 0.40  $\pm$ 6.90 $\times 10^{-4 }$ &      & 0.46  $\pm$ 0.10 & -0.90 $\pm$ 0.19 & 4.92 $\times 10^{-5 }$ & 2.10 & 0.20 \\
          & 4535 & 0.77  $\pm$ 4.00 $\times 10^{-14}$ &      & 0.93  $\pm$ 0.08 & -1.60 $\pm$ 0.15 & 4.44 $\times 10^{-19}$ & 0.73 & 0.05 \\
          & 7496 & 0.54  $\pm$ 6.50 $\times 10^{-5 }$ &      & 0.79  $\pm$ 0.15 & -1.30 $\pm$ 0.25 & 2.03 $\times 10^{-6 }$ & -0.77 & 0.24 \\
 \sigstar & 1566 & 0.85  $\pm$ 3.40 $\times 10^{-31}$ & 0.52 & 1.00  $\pm$ 0.05 & -2.20 $\pm$ 0.10 & 1.48 $\times 10^{-40}$ & 1.50 & 0.08 \\
          & 1672 & 0.50  $\pm$ 9.50 $\times 10^{-7 }$ &      & 0.60  $\pm$ 0.09 & -1.20 $\pm$ 0.17 & 2.52 $\times 10^{-9 }$ & 1.20 & 0.15 \\
          & 3511 & 0.72  $\pm$ 2.60 $\times 10^{-6 }$ &      & 0.59  $\pm$ 0.09 & -1.50 $\pm$ 0.13 & 8.92 $\times 10^{-8 }$ & 2.30 & 0.10 \\
          & 5068 & -0.18 $\pm$ 0.10                   &      & -0.13 $\pm$ 0.15 & -0.01 $\pm$ 0.21 & 0.37                  & 2.70 & 0.11 \\
          & 1954 & 0.10  $\pm$ 0.18                   &      & 0.07  $\pm$ 0.14 & -0.40 $\pm$ 0.22 & 0.72                  & 2.30 & 0.17 \\
 \hline 
                          & 1512 & 0.48  $\pm$ 0.19                    &        & 3.60  $\pm$ 0.63  & -4.10 $\pm$ 0.72 & 1.86 $\times 10^{-7 }$ & -0.06 & 0.18 \\
                          & 4535 & 0.65  $\pm$ 5.20 $\times 10^{-5}$   &        & 2.40  $\pm$ 0.28  & -2.30 $\pm$ 0.29 & 3.70 $\times 10^{-13}$ & -0.12 & 0.07 \\
                          & 7496 & 0.48  $\pm$ 0.01                    &        & 2.00  $\pm$ 0.44  & -1.60 $\pm$ 0.35 & 2.13 $\times 10^{-5 }$ & -0.74 & 0.26 \\
 \zprime($r_{\rm gal}$)   & 1566 & 0.71  $\pm$ 5.90 $\times 10^{-10}$  & 0.48 & 3.60  $\pm$ 0.29  & -3.90 $\pm$ 0.31 & 4.32 $\times 10^{-24}$ & -0.35 & 0.14 \\
                          & 1672 & 0.44  $\pm$ 6.70 $\times 10^{-3 }$  &        & 2.00  $\pm$ 0.35  & -2.30 $\pm$ 0.39 & 1.68 $\times 10^{-7 }$ & -0.10 & 0.16 \\
                          & 3511 & 0.74  $\pm$ 8.30 $\times 10^{-3 }$  &        & 2.00  $\pm$ 0.28  & -2.30 $\pm$ 0.23 & 1.98 $\times 10^{-8 }$ & 0.45 & 0.09 \\
                          & 5068 & -0.12 $\pm$ 0.28                    &        & -0.34 $\pm$ 0.60  & 0.05  $\pm$ 0.42 & 0.55                   & 0.61 & 0.11 \\
                          & 1954 & 0.25  $\pm$ 0.29                    &        & 0.84  $\pm$ 0.66  & -0.94 $\pm$ 0.52 & 0.25                   & 0.33 & 0.16 \\
 \hline 
        & 1512 & 0.68 $\pm$ 0.25 & & 0.65 $\pm$ 0.10 & -2.40 $\pm$ 0.36 & 3.50 $\times 10^{-9 }$ & 6.90 & 0.10 \\
        & 4535 & 0.70 $\pm$ 0.28 & & 0.61 $\pm$ 0.07 & -2.30 $\pm$ 0.30 & 1.04 $\times 10^{-12}$ & 5.90 & 0.06 \\
        & 7496 & 0.91 $\pm$ 0.25 & & 1.30 $\pm$ 0.10 & -5.60 $\pm$ 0.43 & 6.32 $\times 10^{-19}$ & 4.50 & 0.08 \\
\Pdyn   & 1566 & 0.91 $\pm$ 0.24 & 0.89 & 0.90 $\pm$ 0.04 & -4.10 $\pm$ 0.17 & 1.25 $\times 10^{-48}$ & 9.00 & 0.04 \\
        & 1672 & 0.89 $\pm$ 0.12 & & 0.68 $\pm$ 0.04 & -3.00 $\pm$ 0.17 & 2.09 $\times 10^{-31}$ & 13.00 & 0.04 \\
        & 3511 & 0.89 $\pm$ 0.32 & & 1.20 $\pm$ 0.12 & -5.60 $\pm$ 0.51 & 1.48 $\times 10^{-12}$ & 10.00 & 0.04 \\
        & 5068 & 0.33 $\pm$ 0.25 & & 0.30 $\pm$ 0.26 & -1.60 $\pm$ 1.10 & 0.12                   & 11.00 & 0.05 \\
        & 1954 & 0.95 $\pm$ 0.30 & & 0.95 $\pm$ 0.10 & -4.40 $\pm$ 0.41 & 1.28 $\times  10^{-7}$ & 11.00 & 0.02 \\
 \hline 
               & 1512 & 0.31  $\pm$ 0.07                   &        & 0.55  $\pm$ 0.29 & -3.60  $\pm$ 1.90 & 0.06                  & 16.00 & 0.19 \\
               & 4535 & 0.72  $\pm$ 1.10 $\times 10^{-3}$  &        & 1.30  $\pm$ 0.35 & -8.70  $\pm$ 2.40 & 3.82 $\times 10^{-4}$ & 18.00 & 0.11 \\
               & 7496 & -0.01 $\pm$ 0.06                   &        & -0.02 $\pm$ 0.28 & 0.17   $\pm$ 1.90 & 0.97                  & 7.80 & 0.45 \\
 \Pdyn/\sigsfr\ & 1566 & 0.55  $\pm$ 1.40 $\times 10^{-8}$  &  0.40  & 0.97  $\pm$ 0.15 & -6.70  $\pm$ 1.10 & 3.41 $\times 10^{-9}$ & 18.00 & 0.20 \\
               & 1672 & 0.50  $\pm$ 3.20 $\times 10^{-6}$  &        & 0.71  $\pm$ 0.15 & -4.70  $\pm$ 0.98 & 2.41 $\times 10^{-6}$ & 16.00 & 0.16 \\
               & 3511 & 0.34  $\pm$ 0.07                   &        & 0.69  $\pm$ 0.45 & -5.20  $\pm$ 3.10 & 0.10                  & 23.00 & 0.13 \\
               & 5068 & 0.46  $\pm$ 0.05                   &        & 0.50  $\pm$ 0.26 & -3.70  $\pm$ 1.80 & 0.03                  & 25.00 & 0.09 \\
               & 1954 & -0.29 $\pm$ 0.08                   &        &-1.00  $\pm$ 0.90 & 6.60   $\pm$ 6.20 & 0.21                  & 22.00 & 0.16 \\
\hline
\end{tabular}
\begin{minipage}{2.0\columnwidth}
        \vspace{1mm}
        {\bf Notes:} 
        (1): The IC only applies for IC~1954. 
        (2): The correlation coefficient $\rho$ for individual fits for for each galaxy using \texttt{linmix}, see \autoref{sec:linmix_error}. We perform a Monte Carlo analysis perturbing the x and y data points to get the uncertainty of $\rho$. 
        (3): The median $\rho$ for all galaxies for each log(x). 
        (4): The slope $\beta$ of the fit, and 
        (5): The intercept $\alpha$ of the fit. 
        (6): The covariance in the form of Cov($\log_{10}(x)$,$\log_{10}(y)$). (7): The regression intrinsic scatter. 
    \end{minipage}
\end{table*}


\section{Discussion}
\label{sec:discussion}

%
%
%

In this section, we discuss the physical drivers determining the balance between \hi\ and H$_2$ in our sample (i.e. the correlations shown in \autoref{fig:07_scalingrelations}). We further examine conditions at \rmol\ = 1, where above or below this transition, either \htwo\ or \hi\ dominates the cold ISM in nearby galaxies. We also investigate potential differences in physical quantities between interacting and non-interacting galaxies.

\subsection{Dynamical Equilibrium Pressure \texorpdfstring{$P_\mathrm{DE}$}~~and \texorpdfstring{$R_\mathrm{mol}$}~} 

\begin{figure*}[ht!]
    \centering
    \includegraphics[width=1.0\textwidth]{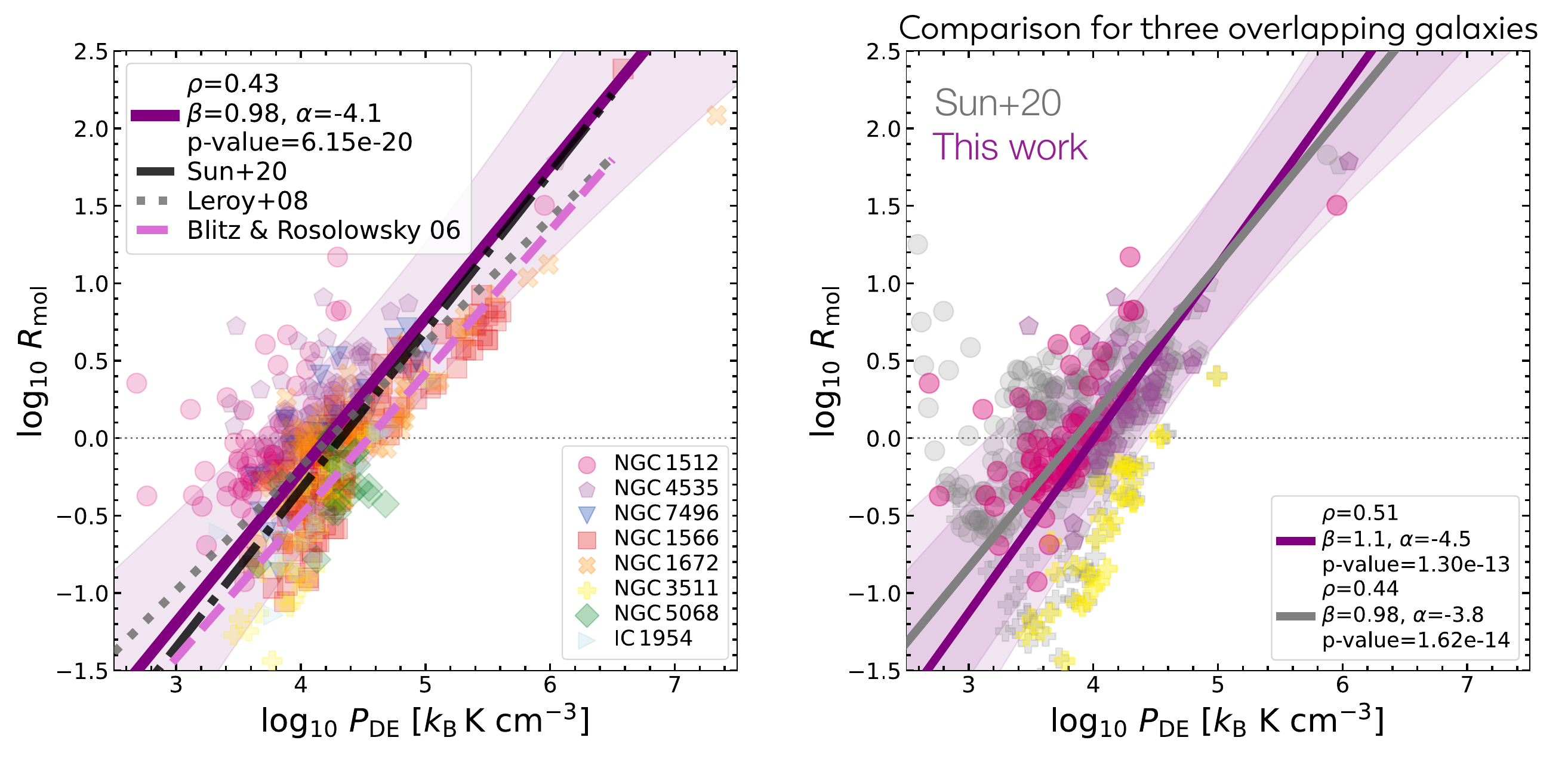}
    \caption[$R_{\rm mol}$ against \Pdyn.]{$R_{\rm mol}$ against \Pdyn. Left: Shown here is an enlarged version of panel 5 from \autoref{fig:07_scalingrelations}, with the fit now performed over the entire sample (purple line) with the corresponding 3$\sigma$ confidence interval (purple shaded region). The black, grey, and pink lines show literature fits adopted from \citet{Sun2020_pde}, \citet{Leroy2008}, and \citet{Blitz2006}, respectively. Right: Comparison between (i) \rmol\ and \Pdyn\ estimates from \cite{Sun2020_pde} for three overlapping galaxies (grey markers) and (ii) our derived \rmol\ and \Pdyn\ (colored markers). The different shapes of the markers correspond to different galaxies.}
    \label{fig:Rmol_vs_PDE}
\end{figure*}

We find a tight, positive correlation between \Pdyn\ and \rmol\ (with $\left<\rho\right>~=0.89$; see panel 5 of \autoref{fig:07_scalingrelations} and \autoref{tab:correlations}), in line with many previous studies.
For example, \citet{Wong2002} reported a strong correlation (with power-law indices of 0.8 and 0.92) between \rmol\ 
and the midplane hydrostatic pressure, mostly in a regime where \sigmol\ > \sigatom. \cite{Blitz2006} found similar results over three orders of magnitude in pressure for 14 galaxies. 
\citet{Leroy2008} showed that this relation also extends to the regime where \hi\ dominates the ISM. However, they stated that the observed correlation between the midplane hydrostatic pressure and \rmol\ could be partly attributed to built-in covariance between the input quantities, as \siggas\ is one of the parameters used to calculate \Pdyn\ (see \autoref{eq:Pde}). Our estimated covariance of the fits in \autoref{tab:correlations} between $\log_{10}$(\rmol) and $\log_{10}$(\Pdyn) are in the range of ${\sim}5-13$, which are the highest among the shown quantities in \autoref{fig:07_scalingrelations} (together with the ones for \sigsfr/\Pdyn).

A study by \citet{Sun2020_pde} measured \Pdyn\ values in the range of $10^{3}-10^{6}$ \PdyKBnunit\ for 28 massive star-forming galaxies similar to our targets. This range is consistent with our \Pdyn\ values of $10^{2.5}-10^{7.5}$\PdyKBnunit, and only three galaxies overlap with their study. \citet{Sun2020_pde} found a moderate positive correlation between \Pdyn\ and \rmol, with Spearman's rank correlation coefficient of ${\rho}=0.58$. With our better quality \hi\ data, we see an even stronger (or tighter) correlation with $\left<\rho\right>~=0.89$ (i.e. the median over all the individual $\rho$ for each galaxy, see \autoref{tab:correlations} and \autoref{fig:07_scalingrelations}). This overall agrees with previous studies \citep{Wong2002,Blitz2006,Leroy2008} modulo different treatments for CO-to-H$_2$ conversion factor, stellar mass-to-light ratio, and stellar disk geometry. 

In the left panel of \autoref{fig:Rmol_vs_PDE} we show \rmol\ against \Pdyn\ for our galaxies together with the aforementioned literature fits \cite[from][]{Blitz2006,Leroy2008,Sun2020_pde}. The solid purple line shows a single ordinary least square (OLS) bisector fit\footnote{We use here and in the following sections the OLS bisector method in logarithmic space \citep[see][for more details]{Isobe1990} to better compare to studies by \cite{Leroy2008} and \cite{Sun2020_pde} who used the same method.} calculated over all of our eight galaxies, which yields a slightly weaker relationship (with $\rho$= 0.43\footnote{The correlation coefficients given here and in the following sections are Spearman rank correlation coefficients, which are suited better for a comparison with the literature than the Pearson correlation coefficient using \texttt{linmix}.}, $\beta$= 0.98, $\alpha$= -4.1, and p-value\,$\ll0.001$) compared to the median over all the individual $\rho$ for each galaxy (i.e. $\left<\rho\right>~=0.89$, see \autoref{tab:correlations}).

In \cite{Sun2020_pde} the \rmol-\Pdyn\ relation was shown for 28 galaxies, using older, less sensitive \hi\ observations (mostly from literature VLA programs) to infer \rmol\ and \Pdyn. Among these 28 galaxies, three overlap with our sample (NGC~1512,  NGC~3511, and NGC~4535), allowing us to directly compare our \rmol\ and \Pdyn\ measurements with theirs. Since we use the same datasets and assumptions for molecular gas and stellar mass components (i.e. \sigmol, $\sigma_{\mathrm{mol}}$, $\rho_\star$ in Equation~\ref{eq:Pde}), we can attribute all differences in \Pdyn\ and \rmol\ to the different \hi\ data entering our analyses.

\begin{figure}[ht!]
    \centering
    \includegraphics[width=0.48\textwidth]{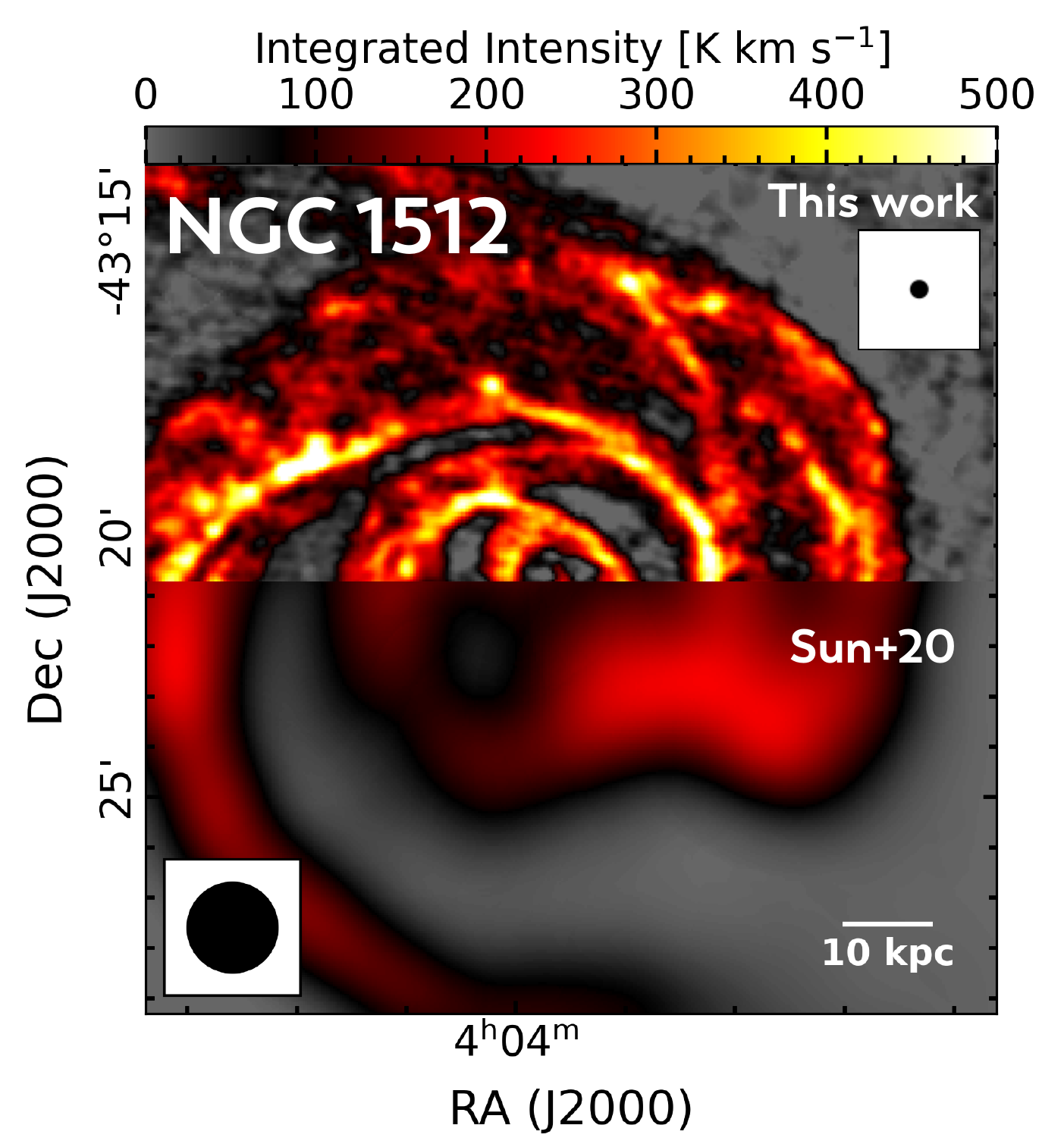}
    \caption{Illustration of different data quality used in \textit{this work} compared to data used in \citealt{Sun2020_pde}. The corresponding beam sizes of ${\sim}15\arcsec$ and ${\sim}100\arcsec$ are visible in the corners.
    }
    \label{fig:ngc1512_new_vs_old}
\end{figure}

In the right panel of \autoref{fig:Rmol_vs_PDE} we show a comparison of the two different estimates of \rmol\ and \Pdyn, for the three overlapping galaxies  NGC~1512,  NGC~3511, and NGC~4535 with \citealt{Sun2020_pde}. We find that our derived \rmol\ (colored markers) are on average smaller than those found in \cite{Sun2020_pde} (grey markers); this is most obvious in the low-pressure regime.  These differences can be attributed to several factors; their older \hi\ observations reached lower sensitivities compared to our MeerKAT observations with an average rms of $\sim 0.3$ mJy/beam over all of our galaxy maps (see \autoref{tab:observations_details}).  More importantly, their VLA data for these three galaxies reached coarser angular resolution of ${\sim}$30\arcsec\ for  NGC~3511, and NGC~4535 but ${\sim}$100\arcsec\ for NGC~1512 compared to our new MeerKAT observations reaching angular resolutions of ${\sim}$15\arcsec. \cite{Sun2020_pde} measured \sigatom\ under the assumption of a smooth sub-beam flux distribution, driven by their limited resolution. With the higher quality data 
we now have available, it is clear that this assumption is not valid and we are now able to measure \sigatom\ at a resolution it is meant to be measured when using 1.5 kpc hexagon regions -- at $<1.5~$kpc (see \autoref{tab:observations_details}). In \autoref{fig:ngc1512_new_vs_old} we show a comparison of both of these datasets for NGC~1512. We find higher integrated intensities, thus higher \sigatom\ in the new observation, that causes \rmol\ to be lower than the \rmol\ measurements in \cite{Sun2020_pde}. This causes the purple fit in the right panel of \autoref{fig:Rmol_vs_PDE} to be steeper than the one performed using the observations from \cite{Sun2020_pde}. Given our comparisons of the three overlapping galaxies, we speculate that the fit by \cite{Sun2020_pde} might be even steeper with high-quality \hi\ data.

\subsection{Star formation surface density \texorpdfstring{$\Sigma_{\rm SFR}$}~versus \texorpdfstring{$R_{\rm mol}$}~}
\label{sec:disc_sfr}

The H$_2$ mass fraction is linked to star formation as reflected by the Kennicutt-Schmidt (KS) relation \citep[][]{Schmidt1959,Kennicutt1998}. Panel 2 of \autoref{fig:07_scalingrelations} shows a moderate positive correlation of \rmol\ with the star formation rate surface density \sigsfr\ (with $\left<\rho\right>~=0.67$). We find covariances in the range of $1.90-4.90$, slightly higher than for \sigstar\ or \zprime($r_{\rm gal}$). 

\begin{figure}[ht]
    \centering
    \includegraphics[width=0.48\textwidth]{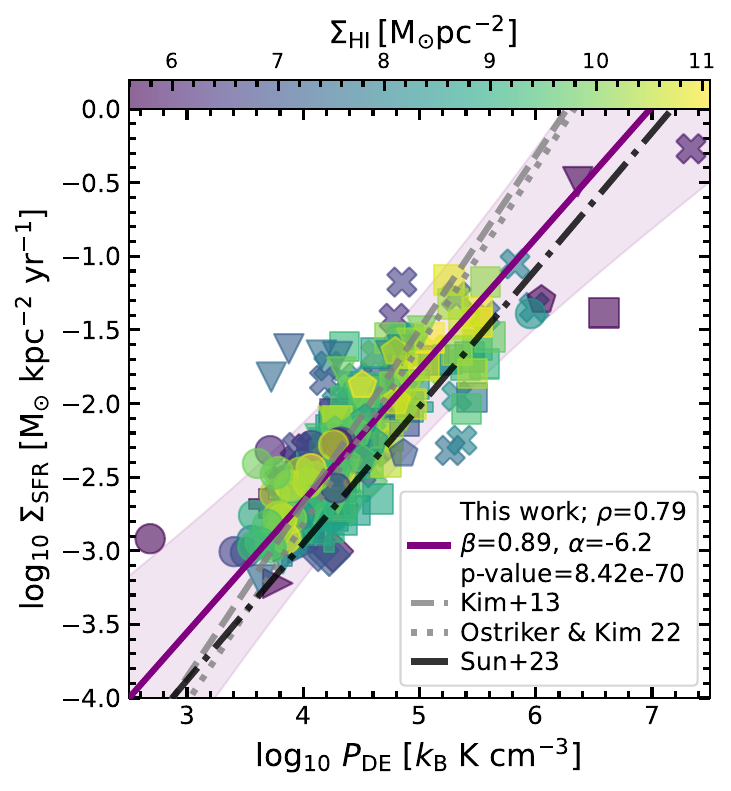}
    \caption{\sigsfr\ against \Pdyn\, colored by \sigatom. The purple line shows our fit over the entire sample with the corresponding 3$\sigma$ confidence interval. The grey dashed and dotted lines show empirical scaling relations from \cite{Kim2013} and \cite{Ostriker2022}. The black dash-dotted line shows a fit from \cite{Sun2022}. We see that low \sigatom\ is responsible for our shallower fit compared to the literature. 
    }
    \label{fig:sfr_vs_pdn}
\end{figure}

The observed correlation between \sigsfr\ and \rmol\ can be understood through their mutual correlation with \Pdyn. We discuss the \Pdyn\ vs \rmol\ relation in the previous subsection; here we show an updated \Pdyn\ vs \sigsfr\ relation (incorporating our new, high-quality \hi\ data) in \autoref{fig:sfr_vs_pdn}. Our observations support that \Pdyn\ plays an important role in setting star formation. An additional factor is that in our sample \hi\ surface densities lower than 8 \uSig\ contribute to a higher intrinsic scatter and shallower fit compared to the indicated literature fits. The gray dashed and dotted lines indicate the empirical scaling obtained from numerical simulations by \cite{Kim2013} and \cite{Ostriker2022}. The black line is drawn from observations by \cite{Sun2023} using 48 star forming spiral galaxies. If we would consider only regions of \sigatom\ > 8 \uSig, then our fit would become as steep as the one shown in \cite{Kim2013} and \cite{Ostriker2022}. Recently it has been shown that galaxies with an offset to the star formation main sequence (SFMS) exhibit a non-linear scaling relation between \sigsfr\ and \Pdyn \citep[see][]{Ellison2024}. Our and the \cite{Sun2023} galaxies are distributed along the SFMS and for those this pressure-regulated star formation theory seems to hold. 

We also examine whether there are galaxy-to-galaxy variations across the \sigsfr-\Pdyn\ plane, and cross-check whether other quantities in our \Pdyn\ estimation could contribute to the shallower relation (e.g \sigmol\ or \siggas). We are unable to find significant galaxy-to-galaxy variations (see \autoref{fig:app:sigsfr_pde}). However, we do notice that the best-fit relation is shallow primarily due to a few low \sigatom\ measurements at $\Pdyn<10^3\,\Pdynunit$ and $>10^6\,\Pdynunit$. The three low \sigatom\ measurements in the $\Pdyn>10^6\,\Pdynunit$ regime are the central regions of NGC~1566, NGC~1672, and NGC~7496, where their AGN is responsible for a \hi-depression. If we were to exclude these few data points, the majority of measurements are broadly consistent with steeper slopes.

\subsection{Other quantities and \texorpdfstring{$R_{\rm mol}$}~}
\label{sec:other_quant_outliers}

In the past, it has been shown that the \hi-to-H$_2$ transition depends on the total gas surface density and metallicity in a PDR \citep[e.g.][]{Sternberg1988}, assuming a spherical cloud that is surrounded by a uniform and isotropic interstellar radiation field \citep[e.g.][]{Krumholz2009,Krumholz2010}. \cite{Krumholz2010} studied the steady-state equilibration and photodissociation of H$_2$ and assumed the \hi-to-H$_2$ transition to be infinitely sharp and suggested that in a given ISM environment (i.e. at fixed metallicity) the total gas surface density determines the transition of gas phases (also tested with observations e.g. \citealt{Fumagalli2010,Lee2012}; or compared with 1D semi-analytical models e.g. \citealt{Sternberg2014}). Panel 1 and panel 4 in \autoref{fig:07_scalingrelations} show that the observationally-derived \rmol\ does moderately correlate with the total gas surface density \siggas\ (with $\left<\rho\right>~=0.67$) and show a weak to moderate correlation with metallicity \zprime($r_{\rm gal}$) (with $\left<\rho\right>~=0.48$). The covariances of each of the galaxies (shown in \autoref{tab:correlations}) of these fits are between $-0.74$ and $0.90$, which indicates that the relationship is not strongly pronounced. 

Looking at individual galaxies, we find `outliers' that are responsible for the wide ranges of $\rho$ values for \siggas, \sigsfr, \sigstar, and \zprime($r_{\rm gal}$). However, there is no one specific galaxy that appears to be an outlier in \textit{all} of the aforementioned quantities. NGC~1512 is responsible for the wide ranges of $\rho$ values for \siggas, that is  \hi-dominated at low \siggas\ regimes. NGC~5068 seems to show different slopes and exhibits negative $\rho$ values in \rmol\ versus \sigsfr, \sigstar, and \zprime($r_{\rm gal}$). This can be attributed to the relatively weak CO emission from this target and consequently the low S/N of the PHANGS-ALMA CO map. This low S/N could introduce biases in measuring \rmol\ and could explain the different behavior seen in the fits. Additionally, the CO map only covers a fraction of the optical disk of NGC~5068, which highlights that additional observations are needed to fully understand this behavior. 
IC~1954 shows a shallow fit between \rmol\ and \sigstar\ because of significant \rmol\ measurements in the low \sigstar\ regime. These two galaxies (NGC~5068 and IC~1954) are also those with the lowest total stellar mass ($M_{\star}$) in our sample (see \autoref{tab:sample_properties}).
We expect an increase in \rmol\ with $\bar{n}_{\mathrm{cloud}}/G_0$. Panel 6 in \autoref{fig:07_scalingrelations} shows globally this behavior for most of our galaxies. However, three galaxies, NGC~7496, NGC~5068, and IC~1954, show the opposite trend. We note that the galaxies NGC~5068 and IC~1954 are according to \autoref{fig:Rmol_vs_r25}  \hi-dominated. However, NGC~7496 is quite the contrary. We can not attribute a specific reason why this galaxy shows this opposite trend. However, the overall trend of increasing \sigmol\ with $\bar{n}_{\mathrm{cloud}}/G_0$ holds for our sample of star forming spiral galaxies. 

\subsection{Differences between interacting and non-interacting galaxies?} 

\autoref{fig:ngc1512} shows a prominent example of an interacting galaxy -- NGC~1512 interacting with NGC~1510. In our sample, we have another galaxy that is potentially interacting with a close-by companion: NGC~3511 with NGC~3513 (see \autoref{fig:07_sample}). Apart from these two galaxies none of the others appear to be interacting\footnote{We note that there are three other galaxies in the field of view of NGC~7496, see \autoref{appendix:additinal_galaxies}. In this paper, we consider them to be non-interacting galaxies, as they show no signs of bridges or tails indicating interaction. However, this may also be a matter of sensitivity, where we are unable to capture this highly diffuse \hi\ gas. Throughout the discussion, we have kept an eye on NGC~7496 to see whether or not it follows a particular trend associated with the two interacting galaxies in our sample. Additionally, we note that NGC~4535 is a member of the Virgo Cluster of galaxies.}. Gas can generally be stripped from a companion, or it can flow inwards because of the torque exerted by the companion or the asymmetric structure created by the interaction. The exchange of gas in different phases can increase the reservoir of molecular gas and replenish the atomic gas by cooling \citep[see e.g.][]{Moreno2019}. 

\cite{Blitz2006} observed a different behavior in \rmol\ vs. pressure between interacting and non-interacting galaxies. From a global view, \citet[][]{Lisenfeld2019} found that mergers have a larger molecular mass gas fraction than their isolated counterparts, resulting in accelerated atomic to molecular gas conversion \citep[][]{Kaneko2017}. More recent studies show evidence that environmental processes, potentially due to interactions, are linked to quenching in \hi-poor galaxies. This simultaneously reduces the molecular gas surface density and star formation efficiency compared to regions in \hi-normal systems \citep[see e.g.][]{Villanueva2022,Jimenez-Donaire2023,Brown2023}.  On the contrary, in galaxy clusters environmental processes can lead to an increase in \rmol\ \citep[see e.g.][]{Moretti2023}. The question arises now whether we find different signatures in \Pdyn, in \rmol, or other physical quantities in these two galaxies compared to the other six. 

What both interacting galaxies have in common are their relatively low global values in star formation rate ${\sim}$0.1~log(M$_{\odot}$yr$^{-1}$) (see \autoref{tab:sample_properties}) among our sample. In the right panel of \autoref{fig:Rmol_vs_r25}, which shows \rmol\ versus $r_{\rm gal}$/$r_{25}$ for individual galaxies, it is evident that NGC~1512  (pink markers and line) shows the smallest radius of where molecular gas becomes dominant $r_{\rm gal}$ = 0.23~$r_{25}$. However, NGC~3511 (yellow markers and line) never crosses the $\rmol=1$ equilibrium and, hence is  \hi-dominated. The first panel in \autoref{fig:07_scalingrelations} and their corresponding intrinsic scatter values listed in \autoref{tab:correlations}, show that NGC~1512 and NGC~3511 have the highest intrinsic scatter in our sample, namely 0.23 and 0.10~dex, respectively. The second panel in \autoref{fig:07_scalingrelations} -- \rmol\ against \sigsfr\ --  does not show a particularly different behavior for NGC~1512 and NGC~3511 as compared to the other 
galaxies in our sample. This also holds for the physical quantities \sigstar, \zprime, and \sigsfr\/\Pdyn. The fifth panel in \autoref{fig:07_scalingrelations} shows that measurements of \rmol\ against \Pdyn\ for NGC~1512 have the highest intrinsic scatter in our sample of 0.10~dex (also see \autoref{fig:app:sigsfr_pde}). However, our sample of a total of eight galaxies, of which only two are interacting, is not statistically significant, and so it is unclear whether galaxy interactions meaningfully impact the \rmol\ relations we are investigating.

\subsection{Atomic vs Molecular dominated ISM}

In spiral galaxies, the transition between a  \hi-dominated and a predominantly  \htwo-dominated ISM takes place at a characteristic value for most quantities. In \autoref{tab:conditions_at_H2-HI} we list the conditions at \rmol\ = 1 and compare our measured values to the ones in \cite{Leroy2008}. 
For each galaxy, we measured the median of the quantity in question over all hexagonal apertures where \sigmol $\approx$ 0.8-1.2~\sigatom. We show the median transition value along with the 1$\sigma$ scatter among our galaxies. At $\rmol= 1$ the baryon mass budget is dominated by stars \sigstar $\approx78$~\uSig~ while the total gas surface density is \siggas $\approx14$~\uSig. The dynamical equilibrium pressure corresponds to 2.9$\times10^4$\,\PdyKBnunit. We note that some of these values are different from those quoted in \autoref{tab:conditions_at_H2-HI} taken from \cite{Leroy2008}, but the mean values overlap within the scatter, and since these quantities depend on the galaxy's properties, a perfect match is not necessarily to be expected.

In the right panel of \autoref{fig:Rmol_vs_r25} and in \autoref{tab:07_rmol_r25} we found that three galaxies never cross the \rmol\ equilibrium border to enter the molecular gas dominated regime (i.e. \rmol\ > 1). These  \hi-dominated galaxies are NGC~3511, NGC~5068, and IC~1954. NGC~5068, and IC~1954 are the lowest $M_{\star}$ galaxies in our sample with $M_{\star}$ = 9.4 log(M$_{\odot}$) and $M_{\star}$ = 9.7 log(M$_{\odot}$), respectively. We note that this shows that it is important to study each galaxy individually, as they show different behaviors in \rmol\ under different local conditions (see also \autoref{sec:other_quant_outliers}). However, we also believe that listing the scaling relation across all galaxies, similar to what we did for \rmol\ versus \rgal/$r_{25}$ in \autoref{fig:Rmol_vs_r25} and \autoref{eq:rmol_r25}, may be useful for future comparisons, and therefore we list them in \autoref{tab:scaling_relations_across_all}.

\begin{table}
    \centering
    \caption[Conditions at $R_{\rm mol} = 1$, where \sigatom$\approx$\sigmol]{Conditions at $R_{\rm mol} = 1$, where \sigatom$\approx$\sigmol.}
    \begin{tabular}{lccc}
    \hline \hline
    Quantity & Median  & Scatter & Leroy+08 \\
     & (1)  & (2) & (3) \\
    \hline
   $r_{\rm gal}$/$r_{25}$  & 0.36 & 0.16 & 0.43 \\ 
      \siggas~[\uSig] & 14 & 2 & 14\\ 
      \sigsfr~[\sigsfrunit] & 3.0$\times10^{-3}$ & 2.3$\times10^{-3}$ &  - \\ 
      \sigstar~[\uSig] & 78 & 37 & 81\\ 
      \Pdyn~[\PdyKBnunit] & 3.4$\times10^4$ & 4.4$\times10^3$ & 2.3$\times10^4$\\ 
        \hline
    \end{tabular}
    \label{tab:conditions_at_H2-HI}
    \begin{minipage}{1.0\columnwidth}
        \vspace{1mm}
        {\bf Notes:} 
        (1): Following \cite{Leroy2008} we measure the quantity for each galaxy over all apertures where \sigmol = $0.8-1.2$~\sigatom, then take the median. (2): The corresponding 1$\sigma$ scatter. (3): Literature median values using observations from the THINGS and HERACLES surveys \citep[][]{Leroy2008}.
    \end{minipage}  
\end{table}

\begin{table}[]
    \centering
    \caption{Scaling relation across all galaxies.}
    \begin{tabular}{lc}
    \hline \hline
     & Figure \\
    \hline
    $\log_{10} \rmol = -0.9 + 0.73 \times \log_{10} \left(\frac{\siggas}{\uSig}\right)$ &  p1 \\
    $\log_{10} \rmol = 1.7 + 0.75 \times \log_{10} \left(\frac{\sigsfr}{\sigsfrunit}\right) $   &  p2 \\
    $\log_{10} \rmol = -1.2 + 0.64 \times \log_{10} \left(\frac{\sigstar}{\uSig}\right)$ &   p3 \\
    $\log_{10} \rmol = -1.3 + 1.2 \times \zprime(r_{\rm gal})$ & p4\\
    $\log_{10} \rmol = -2.1 + 0.49 \times \log_{10} \left(\frac{\Pdyn}{\PdyKBnunit}\right)$ & p5\\
    $\log_{10} \rmol = -2.4 + 0.36 \times \log_{10} \left(\frac{\Pdyn}{\sigsfr}\right)$ & p6\\
    $\log_{10} \left(\frac{\sigsfr}{\sigsfrunit}\right) =  -6.2 +  0.89 \times \log_{10} \left(\frac{\Pdyn}{\PdyKBnunit}\right)$ & F11\\
    \hline 
    \end{tabular}
    \label{tab:scaling_relations_across_all}
    \begin{minipage}{1.0\columnwidth}
        \vspace{1mm}
        \textbf{Notes:}  In the form of $y=\alpha+\beta~\times~x$. p1-p6 refers to panel 1 to panel 6 in \autoref{fig:07_scalingrelations}. F11 refers to \autoref{fig:sfr_vs_pdn}.
    \end{minipage}
\end{table}


\section{Conclusion}
\label{sec:conclusion}

We used a multi-wavelength approach including new observations of eight nearby galaxies taken with the MeerKAT telescope (PHANGS-MeerKAT, this work + MHONGOOSE from \citealt{deBlok2016}). In this work, we analyzed at what galactocentric radii atomic gas becomes more dominant compared to molecular gas and examined local conditions in individual galaxies and how they correlate with the molecular gas fraction \rmol\ = \sigmol/\sigatom. We performed our analysis in a way that we averaged each quantity (\sigmol, \sigatom,  \siggas, \sigsfr, \sigstar, \zprime($r_{\rm gal}$), \Pdyn, and \sigsfr\ / \Pdyn) over 1.5~kpc hexagonal regions (following \citealt{Sun2022}). We find the following results:

\begin{itemize}
    \item[1)] We presented new MeerKAT \hi\ observations of three nearby spiral galaxies NGC~1512, NGC~4535, and NGC~7496 (PHANGS-MeerKAT) at 15\arcsec angular resolution, which we combined with additional observations towards five nearby galaxies \samplemhongoose\ (from MHONGOOSE); see \autoref{fig:ngc1512_color} -- \autoref{fig:07_sample}. Their \hi\ distribution shows gigantic spiral arms (in NGC~1512, NGC~1566, and NGC~4535), splitting into sub-arms (in NGC~1512), compact disks (in NGC~3511 and IC~1954), ring-like structures (in NGC~7496 and a smaller one in NGC~1512), or additional clumps (in NGC~1512 and NGC~1566).    
    
    \item[2)] We detected on average significant \hi\ emission out to a radius of ${\sim}50$~kpc (see \autoref{fig:07_radial_extent_ngc1512}, \autoref{fig:07_radial_extent_big}, and \autoref{fig:07_radial_extent_small}), which is ${\sim}$5 times the 3$\sigma$ detected CO emission (see \autoref{tab:07_r25}). In terms of $r_{\hi}$, defined as the \sigatom\ = 1~\uSig\ isophote, we found an average $r_{\mathrm{HI}}$ of  ${\sim}22~$kpc, which is ${\sim}2~$ times $r_{25}$ (with ranges of  $1-3$ times $r_{25}$) and  ${\approx}6$ times $r_{\rm eff}$.
    
    \item[3)] The \rmol\ - $r_{\rm gal}/r_{25}$ fit crosses the \rmol\ = 1 border (where \sigatom $\approx$ \sigmol, indicated as a grey horizontal line in \autoref{fig:Rmol_vs_r25}) at $r_{\rm gal}~{\sim}0.4~r_{25}$. This means that molecular gas is more dominant at $0{-}0.4$ optical radii, and atomic gas becomes more dominant at $r_{\rm gal}~>0.4$ optical radii. 

    \item[3)] 
    To understand how \rmol\ responds to the local conditions in each galaxy, we investigated scaling relations with various physical parameters using a hierarchical Bayesian method to include uncertainties and upper limits (see \autoref{sec:linmix_error}). Among all of these fits \rmol\ - \Pdyn\ shows the tightest correlation with a median correlation coefficient $\rho$ of each galaxy's $\rho$ of $\left<\rho\right>=0.89$, together with the smallest intrinsic scatter (${\sim}0.1$dex; see \autoref{fig:07_scalingrelations} and \autoref{tab:correlations}).

    \item[4)] A positive correlation between \rmol\ and \Pdyn\ has been found in previous studies \citep[e.g.][]{Blitz2004,Leroy2008,Sun2020}. Similar to these studies we performed a OLS bisector fit over all galaxies, which yields for our sample a correlation with $\rho = 0.43$ (see \autoref{fig:Rmol_vs_PDE}). We attribute the differences in our \rmol\ measurements compared to previous studies to the improved resolution and sensitivity of our \hi\ data. Upcoming observations will increase the sample size and thus allow for further investigation. 

    \item[5)] We investigated how the dynamical equilibrium pressure corresponds to the star formation rate. We found that \sigatom\ measurements  lower than 8 \uSig\ are responsible for a shallower fit (see \autoref{fig:sfr_vs_pdn}) compared to the empirical scaling relations from \cite{Kim2013,Ostriker2022}  but agree with the observational one by \cite{Sun2023}. 

    \item[6)] In our sample, NGC~1512 interacts with NGC~1510 and potentially NGC~3511 with its close-by companion NGC~3513. We did not find a significantly different behavior than the non- interacting galaxies in our sample among all of our physical quantities. 

    \item[7)] We found that the transition between an \hi-dominated and \htwo-dominated ISM takes place at a characteristic value for most quantities we measured. For example, at \rmol\ = 1 the baryon mass budget is dominated by stars \sigstar > 5~\siggas  (i.e. \sigstar$\approx78$~\uSig\ while the total gas is \siggas $\approx14$~\uSig; see \autoref{tab:conditions_at_H2-HI}).

\end{itemize}

\label{sec:next_steps}
With upcoming MeerKAT and/or high-resolution VLA \hi\ observations of additional nearby galaxies targeted by the PHANGS and/or MHONGOOSE projects that have multi-wavelength data available\footnote{The availability of CO observations is required to calculate for example \Pdyn, see \autoref{sec:quant_pde}, \autoref{eq:Pde}.}, a natural next step is to perform this analysis on a larger sample. With these fantastic wide-field \hi\ surveys, wider CO mapping would also extend these relations into the outer disk environments (a first step towards this goal has been taken within PHANGS; observing CO(1-0) towards 10 galaxies with ALMA's compact array, PI:~A.~Leroy). This will allow us to further investigate how global galaxy properties (stellar mass, star formation rate, or morphology) impact the conversion from atomic to molecular gas in nearby galaxies ($D < 20$~Mpc or $z < 0.005$).


\begin{acknowledgements}
    CE gratefully acknowledges funding from the Deutsche Forschungsgemeinschaft (DFG) Sachbeihilfe, grant number BI1546/3-1. CE acknowledges funding from the European Research Council (ERC) under the European Union’s Horizon 2020 research and innovation program (grant agreement No.726384/Empire). 

    JS acknowledges support by the National Aeronautics and Space Administration (NASA) through the NASA Hubble Fellowship grant HST-HF2-51544 awarded by the Space Telescope Science Institute (STScI), which is operated by the Association of Universities for Research in Astronomy, Inc., under contract NAS~5-26555.

    The work of AKL is partially supported by the National Science Foundation under Grants No. 1615105, 1615109, and 1653300.

    WJGdB, FMM, and JH acknowledge funding from the European Research Council (ERC) under the European Union’s Horizon 2020 research and innovation programme (grant agreement No. 882793 ``MeerGas'').

    M.C. gratefully acknowledges funding from the DFG through an Emmy Noether Research Group (grant number CH2137/1-1). COOL Research DAO is a Decentralized Autonomous Organization supporting research in astrophysics aimed at uncovering our cosmic origins.

    RSK  acknowledges funding from the European Research Council via the ERC Synergy Grant ``ECOGAL'' (project ID 855130), from the German Excellence Strategy via the Heidelberg Cluster of Excellence (EXC 2181 - 390900948) ``STRUCTURES'', and from the German Ministry for Economic Affairs and Climate Action in project ``MAINN'' (funding ID 50OO2206). He also thanks for computing resources provided by {\em The L\"{a}nd} and DFG through grant INST 35/1134-1 FUGG and for data storage at SDS@hd through grant INST 35/1314-1 FUGG.

    HAP acknowledges support from the National Science and Technology Council of Taiwan under grant 110-2112-M-032-020-MY3.

    AU acknowledges support from the Spanish grants PID2019-108765GB-I00, funded by MCIN/AEI/10.13039/501100011033, and PID2022-138560NB-I00, funded by MCIN/AEI/10.13039/501100011033/FEDER, EU. 
    
    ER acknowledges the support of the Natural Sciences and Engineering Research Council of Canada (NSERC), funding reference number RGPIN-2022-03499.

    DJP and SK greatly acknowledge support from the South African Research Chairs Initiative of the Department of Science and Technology and National Research Foundation.\\

This paper makes use of MeerKAT observations. The MeerKAT telescope is operated by the South African Radio Astronomy Observatory, which is a facility of the National Research Foundation, an agency of the Department of Science and Innovation. \\

This paper makes use of the following ALMA data, which have been processed as part of the PHANGS–ALMA CO(2–1) survey:\\
ADS/JAO.ALMA\#2015.1.00925.S, \linebreak 
ADS/JAO.ALMA\#2015.1.00956.S, \linebreak 
ADS/JAO.ALMA\#2017.1.00392.S, \linebreak 
ADS/JAO.ALMA\#2017.1.00886.L, \linebreak 
ADS/JAO.ALMA\#2018.1.01651.S. \linebreak 
ADS/JAO.ALMA\#2018.A.00062.S. \linebreak 
ALMA is a partnership of ESO (representing its member states), NSF (USA), and NINS (Japan), together with NRC (Canada), MOST and ASIAA (Taiwan), and KASI (Republic of Korea), in cooperation with the Republic of Chile. The Joint ALMA Observatory is operated by ESO, AUI/NRAO, and NAO.\\

 Figure 1 in this paper uses hips2fits,\footnote{https://alasky.cds.unistra.fr/hips-image-services/hips2fits} a service provided by CDS.

\end{acknowledgements}

\bibliographystyle{aa}
\bibliography{references}

\appendix

\section{Star formation rate surface density}
\label{appendix:sfr}

\begin{figure}[ht]
    \centering
    \includegraphics[width=0.48\textwidth]{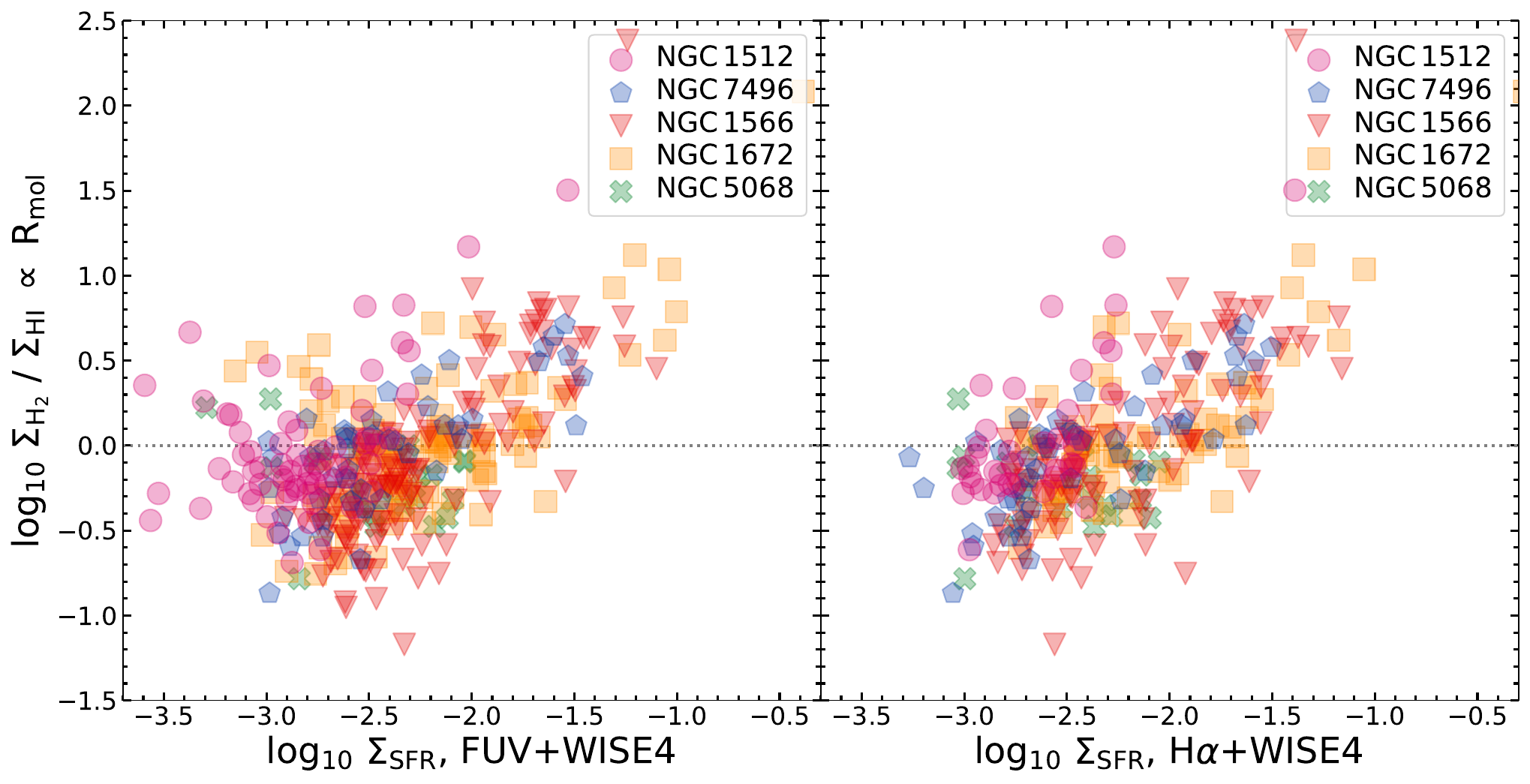}
    \caption{Comparison of star formation rate surface densities using FUV+WISE4-based and the H$\alpha$-based measurements in the \rmol - \sigsfr\ plane. }
    \label{fig:SigSFR_comp}
\end{figure}

In this section, we compare two different methods for computing the star formation rate surface density \sigsfr. We recall that in the main part of this work, we used the maps based on a combination of FUV and mid-IR emission, since these observations are available for all galaxies in our study (see \autoref{sec:quant_sigsfr}). Extinction-corrected H$\alpha$-based SFR measurements together with the new calibration by \cite{Belfiore2023} are considered to be one of the most reliable SFR measurements. 
However, the required MUSE data from the PHANGS-MUSE sample (see \citealt{Emsellem2022}) are available for only 5 out of 8 of the galaxies studied in this work (see \autoref{tab:phangs_muse_coverage}). We examine here the reliability of the \rmol\ versus \sigsfr\ fit (shown in \autoref{fig:07_scalingrelations} second panel) by comparing it to the H$\alpha$-based SFR in the 5 galaxies where both SFR measurements are available. \autoref{fig:SigSFR_comp} shows that the relation using the H$\alpha$-based \sigsfr\ shows a similar trend but with less scatter and thus might be indicative of a tighter relationship between \rmol\ and \sigsfr.  

\begin{table}[h]
\centering
\caption{PHANGS-MUSE coverage of our sample.}
\begin{tabular}{lcc}
\hline \hline
Galaxy &  Project & in PHANGS-MUSE\\
& code &  \\
& (1) & (2)  \\
\hline
IC 1954   & M   & \xmark \\
NGC 1512  & P-M & \cmark \\
NGC 1566  & M   & \cmark \\
NGC 1672  & M   & \cmark\\
NGC 3511  & M   & \xmark \\
NGC 4535  & P-M & \xmark \\
NGC 5068  & M   & \cmark \\
NGC 7496  & P-M & \cmark \\
\hline
\end{tabular}
\begin{minipage}{1.0\columnwidth}
        \vspace{1mm}
        {\bf Notes:} 
        (1): Project codes: 
        P-M = PHANGS-MeerKAT (cycle 0 observation); \textit{this work}. 
        M = MHONGOOSE \citep[see][]{deBlok2016}. 
        (2): PHANGS-MUSE survey \citep[see][]{Emsellem2022,Groves2023}
    \end{minipage}
\label{tab:phangs_muse_coverage}
\end{table}

\section{Metallicity}
\label{appendix:metallicity}

\begin{figure*}[ht]
    \centering
    \includegraphics[width=0.9\textwidth]{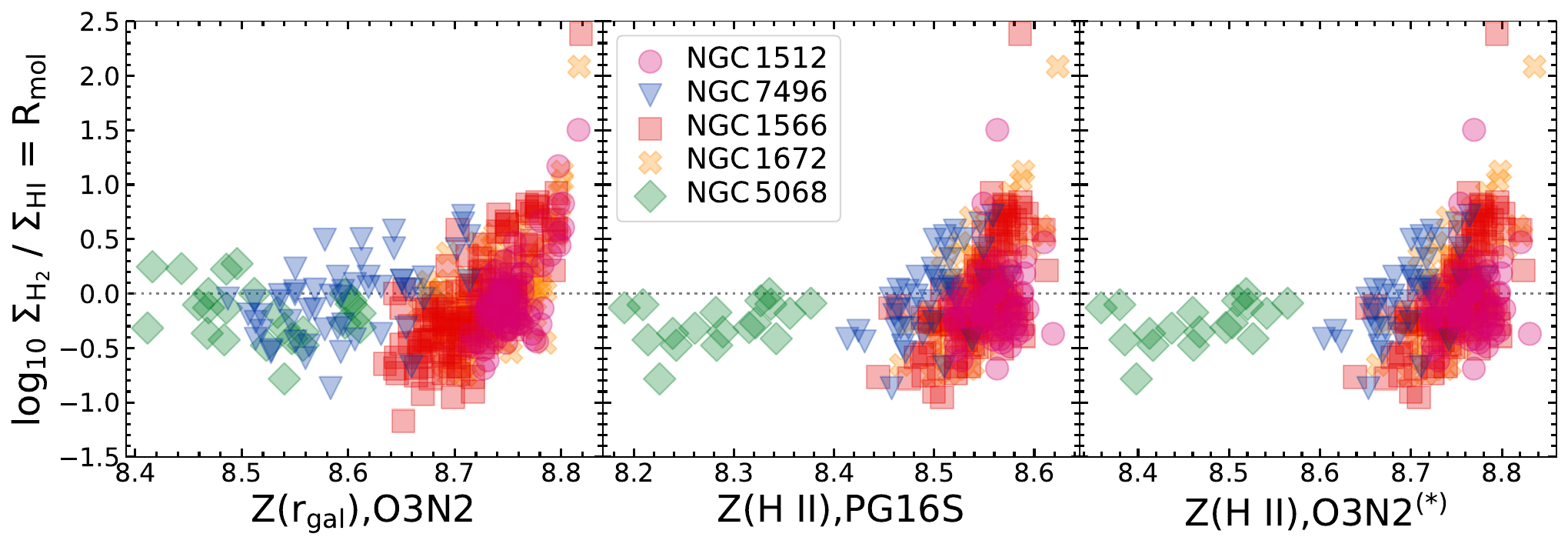}
    \caption{Comparison of metallicity measurements (i.e. 12+log(O/H)). \textit{Left:} Using fixed radial metallicity gradient using the O3N2 calibration (that uses the [OIII]$\lambda$5007, [NII]$\lambda$6584, H$\beta$ and H$\alpha$ lines, see \citealt{Pettini2004}). \textit{Middle:} For individual HII regions using the PG16S calibration (that uses the [OIII]$\lambda$4959,5007, [NII]$\lambda$6584,6584, [SII]$\lambda$6717,6731, H$\beta$ and H$\alpha$ lines, see \citealt{Pilyugin2016}). \textit{Right:} Measurements from the middle panel converted to O3N2 calibration using the third panel Figure B1 in the Appendix of \cite{DeVis2019}.}
    \label{fig:Metallicity_comp}
\end{figure*}

In this section, we compare two different approaches to estimate the metallicity. We recall that in the paper we used scaling relations to get general metallicity trends assuming a global galaxy mass-metallicity relation and a fixed radial metallicity gradient within each galaxy (see \autoref{sec:quant_metal}). In this section, we compare this approach with the use of metallicity measurements of individual HII regions for the galaxies in the PHANGS-MUSE sample (see \autoref{tab:phangs_muse_coverage}). They have been derived in \cite{Groves2023} as the mean HII region metallicity relative to solar using the PG16S calibration (see also \citealt{Kreckel2019}) that is defined in \cite{Pilyugin2016}. This prescription is based on a calibration against a sample of a few hundred HII regions where direct auroral line detections were used to calculate the electron temperature \citep[see][for more details]{Groves2023}. \autoref{fig:Metallicity_comp} shows that comparison (not normalized to the solar metallicity, as we did in the main part of the paper), and it is immediately apparent the majority of galaxies is "moved" to lower metallicity measurements -- ${\sim}0.2$ dex lower $Z$(HII),PG16S compared to $Z$(r$_{\rm gal}$),O3N2. Differences in individual HII region metallicities based on different measurement approaches have been reported to differ up to 0.7~dex in 12+log(O/H) for the same HII regions \citep[][]{Kewley2008}. To check the variations of these two calibration methods we used the third panel of Figure B1 in the Appendix of \cite{DeVis2019} that compares these calibration methods. We use a linear fit of the form of $y=1.1*x-0.65$ and limit to relevant metallicity ranges to "convert" (denoted as $^{(*)}$) our $Z$(HII),PG16S to $Z$(HII),O3N2$^{(*)}$. We find ${\sim}0.2$ dex difference between $Z$(HII),PG16S and $Z$(HII),O3N2$^{(*)}$, consistent with the difference aforementioned. One galaxy --  NGC~5068 (green quared markers) -- is in both measurements distinct from the other galaxies: it is more metal-poor. Neglecting NGC~5068 at \rmol = 1 reveals that $Z$(HII),PG16S$~\approx8.55$ and $Z$(HII),O3N2$^{(*)}$$~\approx8.75$. The galaxy NGC~7496 appears according to the $Z$(r$_{\rm gal}$),O3N2 measurements more metal-poor compared to the $Z$(HII),PG16S and $Z$(HII),O3N2$^{(*)}$ measurements.

Overall both of the calibrations generally yield similar metallicity gradients, just with an offset in the normalization (see e.g. Figure 20 in \cite{Kreckel2019}). Therefore, changing prescriptions just alters $Z$ and \rmol\ by a constant factor (the latter because of the Z-dependence of the CO-H$_{2}$ conversion factor), and hence we do not expect much of an impact on the correlation between \rmol\ and $Z$. 

\section{Assuming a constant \texorpdfstring{$\sigma_{\mathrm{atom}}$}~in \autoref{eq:Pde} }
\label{appendix:pdyn}

\begin{figure}[ht]
    \centering
    \includegraphics[width=0.49\textwidth]{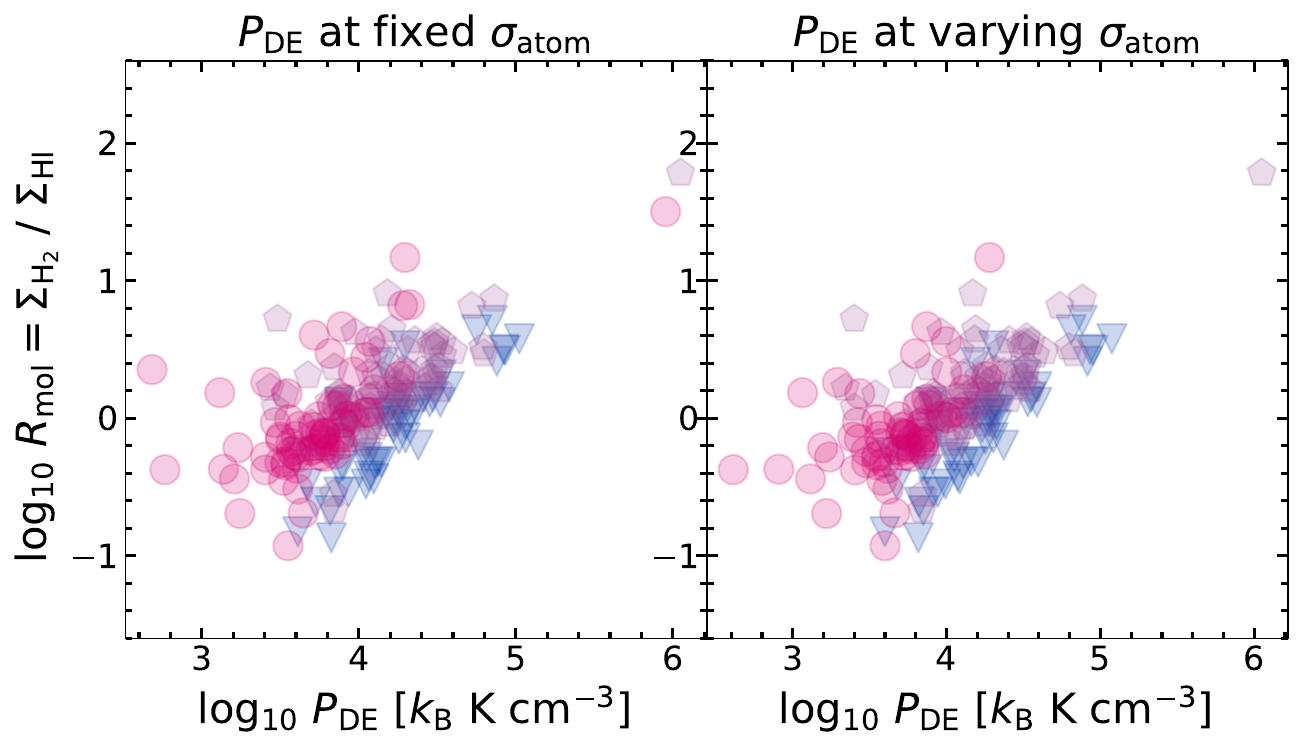}
    \caption{Our \Pdyn\ estimates using a fixed $\sigma_{\mathrm{atom}}$ of 10~\kms\ (left panel) and a varying $\sigma_{\mathrm{atom}}$ (right panel). We only show the comparison for the PHANGS-MeerKAT galaxies, for which $\sigma_{\mathrm{atom}}$ derived from the line Equivalent Width (EW) maps is available (see \citealt{Leroy2021_survey} for details about EW).}
    \label{fig:Pdyn_comp}
\end{figure}

In \autoref{eq:Pde} we fixed $\sigma_{\mathrm{atom}}$ to 10~\kms. It was already highlighted in \cite{Sun2020_pde} that changing $\sigma_{\mathrm{atom}}$ has a minimal impact on the derived \Pdyn\ estimates. In \autoref{fig:Pdyn_comp} we show for the PHANGS-MeerKAT galaxies that \Pdyn\ derived using (i) a fixed $\sigma_{\mathrm{atom}}$, and (ii) a varying $\sigma_{\mathrm{atom}}$, differ only slightly. 

\section{No galaxy-to-galaxy variation across the \texorpdfstring{\sigsfr-\Pdyn\ }~plane}

\begin{figure*}[ht]
    \centering
    \includegraphics[width=1.0\textwidth]{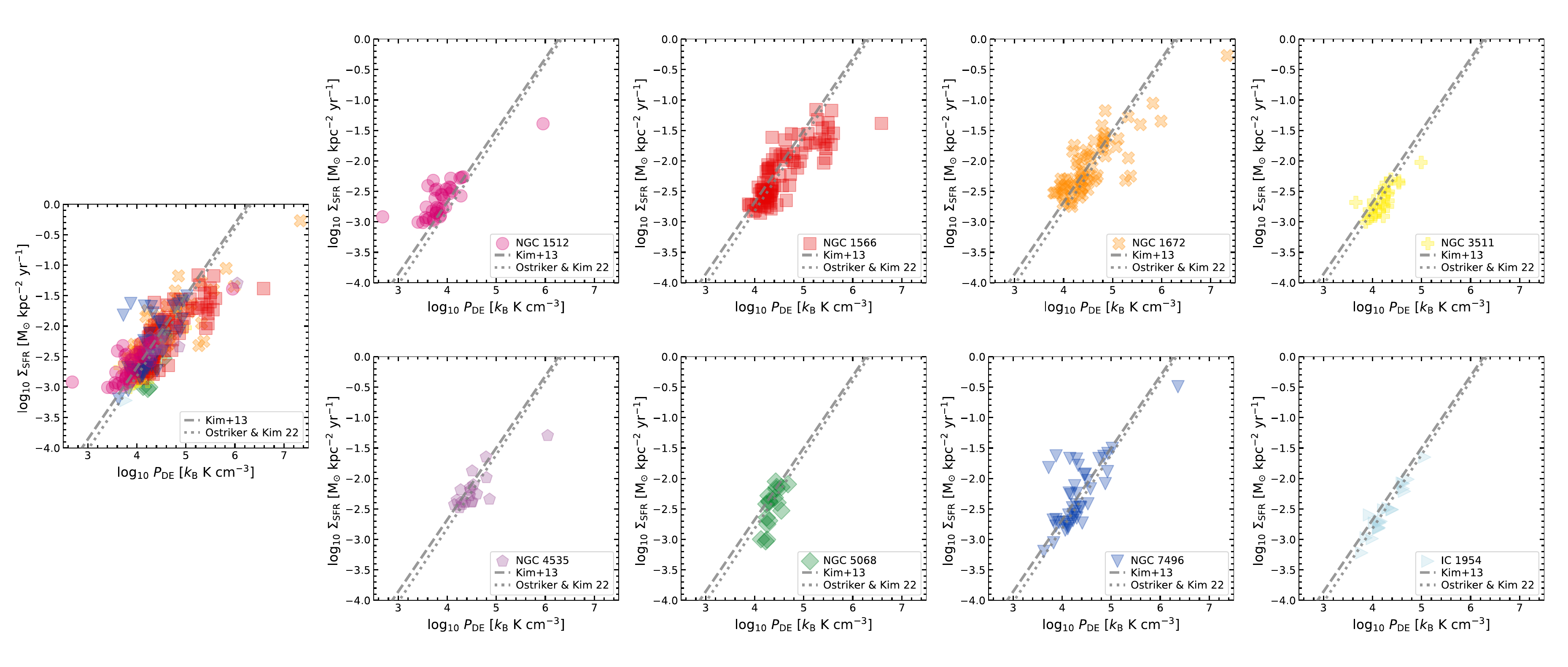}
    \caption{We show here \sigsfr\ against \Pdyn\ for individual galaxies. We do not find any specific or significant variation between galaxy characteristic (e.g. interacting, \hi\ dominant)}
    \label{fig:app:sigsfr_pde}
\end{figure*}

In the \autoref{sec:disc_sfr} we have shown that \Pdyn\ plays an important role in setting star formation. \autoref{fig:app:sigsfr_pde} shows that we are unable to find significant galaxy-to-galaxy variations for \sigsfr\ versus \Pdyn. We show individual measurements for individual galaxies (8 panels to the right) together with the empirical scaling relation (grey dashed lines). We note again the measurements in the $\Pdyn>10^6\,\Pdynunit$ regime that showed up in \autoref{fig:sfr_vs_pdn} as low \sigatom\ measurements which are attributed to the central regions of NGC~1566, NGC~1672, and NGC~7496, where their AGN is responsible for a \hi-depression. 

\section{Additionally detected galaxies in the PHANGS-MeerKAT field of view}
\label{appendix:additinal_galaxies}

\begin{figure}[ht]
    \centering
    \includegraphics[width=0.5\textwidth]{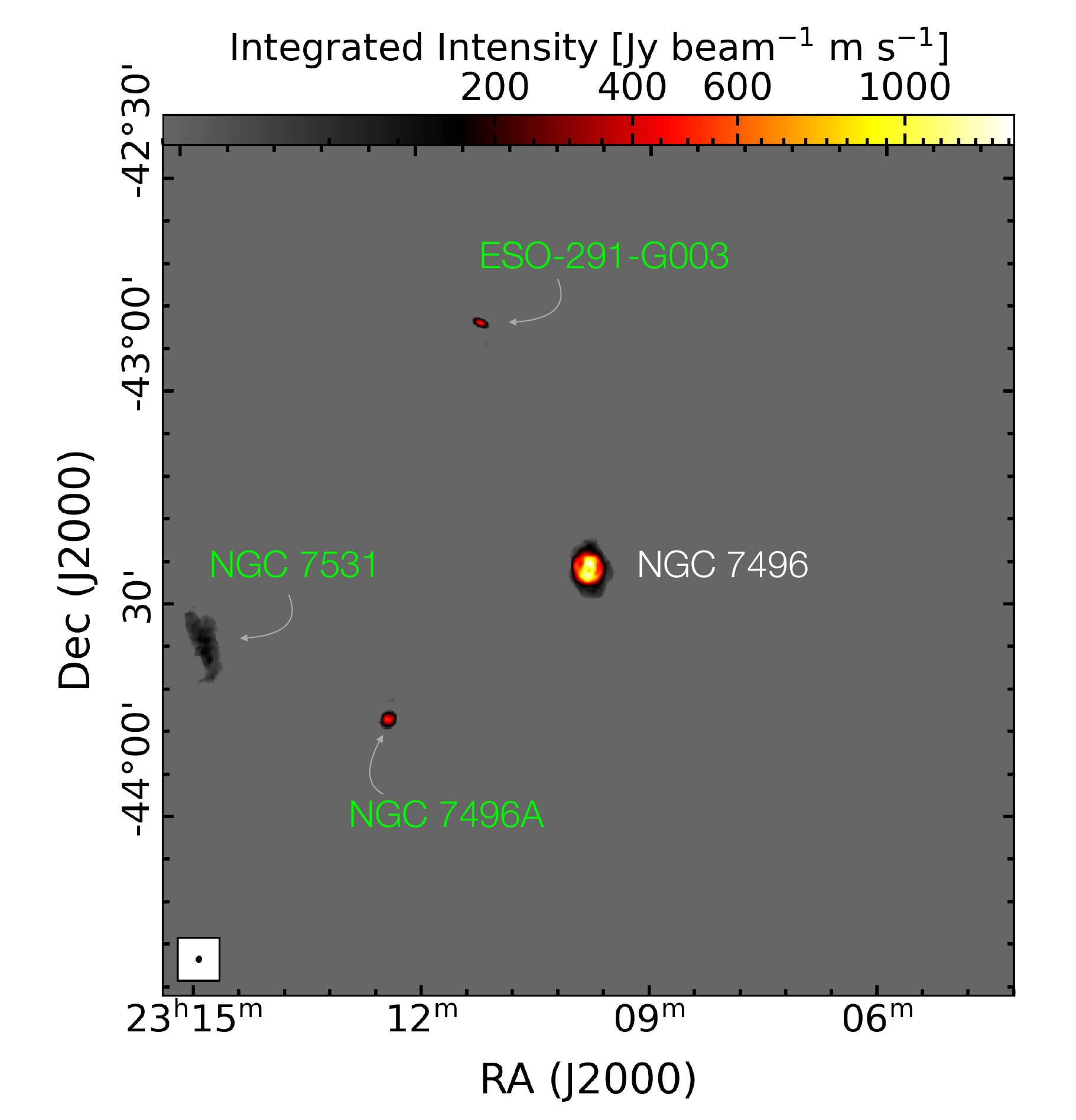}
    \caption{Additionally detected galaxies around NGC~7496.}
    \label{fig:app_mom0}
\end{figure}

We find additional galaxies that were captured within the field of view of MeerKAT. We targeted NGC~7496, but can clearly see in \autoref{fig:app_mom0} three additional galaxies. These are NGC~7496~A, NGC~7531 and ESO~291-G003. The integrated intensity map was imaged at 30\arcsec resolution. In the main part of this paper, however, we focus on galaxies that have all the relevant ancillary data available and imaged them at 15\arcsec resolution.  

\label{LastPage}
\end{document}

%% file: mycommands.tex
\newcommand{\todo}{\ifmmode \text{\color{red}\Huge{\(\bullet\)}} \else {\color{red}{\Huge$\bullet$}}\fi}

\newcommand{\sbsc}[1]{_\mathrm{#1}}
\newcommand{\spsc}[1]{^\mathrm{#1}}

\newcommand{\Ihi}{\ensuremath{I_{_{\mathrm{HI}}}}}
\newcommand{\CO}[2]{\mbox{$\mathrm{CO}\,(#1\text{--}#2)$}}
\newcommand{\hi}{\ifmmode {\rm H}\,\textsc{i} \else H\,\textsc{i}\fi}
\newcommand{\htwo}{{H$_2$}}

\newcommand{\alphaCO}{\alpha\sbsc{CO}}
\newcommand{\alphaCOline}[2]{\alpha\sbsc{CO(#1\text{--}#2)}}
\newcommand{\Rsub}[1]{R_\mathrm{#1}}
\newcommand{\rgal}{$r_\mathrm{gal}$}
\newcommand{\rmol}{\ensuremath{R_{\mathrm{mol}}}}

\newcommand{\sample}{{NGC\,1512, NGC\,4535, and NGC\,7496}}
\newcommand{\samplemhongoose}{IC1954, NGC\,1566, NGC\,1672, NGC\,3511, and NGC\,5068}

\newcommand{\sn}{\ensuremath{\mathrm{S/N}}}
\newcommand{\kms}{{km$\,$s$^{-1}$}}
\newcommand{\uv}{{$u{-}v$\,}}

\newcommand{\cmark}{\textcolor{green!80!black}{\ding{51}}}
\newcommand{\xmark}{\textcolor{red}{\ding{55}}}

\newcommand{\sigatom}{\ensuremath{\mathrm{\Sigma_{HI}}}}

\newcommand{\sigmol}{\ensuremath{\mathrm{\Sigma_{H2}}}}

\newcommand \sigsfr {\ensuremath{{\Sigma_\mathrm{SFR}}}}
\newcommand \sigsfrunit {\ensuremath{\mathrm{M_{\odot}\,yr^{-1}\,kpc^{-2}}}}

\newcommand{\sigstar}{\ensuremath{\mathrm{\Sigma_{*}}}}

\newcommand{\Pdyn}{\ensuremath{P\sbsc{DE}}}
\newcommand{\Pdynunit}{\ensuremath{\mathrm{K\,cm^{-3}}}}
\newcommand{\PdyKBnunit}{\ensuremath{k_{\mathrm{B}}\mathrm{\,K\,cm^{-3}}}}

\newcommand{\siggas}{\ensuremath{\mathrm{\Sigma_{gas}}}}
\newcommand{\sigtot}{\ensuremath{\mathrm{\Sigma_{HI}+\Sigma_{H_2}}}}

\newcommand{\zprime}{\ensuremath{Z'}}

\newcommand{\Mstar}{M_\star}

\newcommand{\Reff}{r\sbsc{e}}

\newcommand{\ICO}{I\sbsc{CO}}

\newcommand{\Iiracone}{I\sbsc{3.6\,\mu m}}
\newcommand{\MtoLwiseone}{\Upsilon_{3.4\,\mu m}}

\newcommand{\uI}{\mbox{$\rm MJy~sr^{-1}$}}
\newcommand{\uIco}{\mbox{$\rm K~km~s^{-1}$}}

\newcommand{\uIhi}{\mbox{$\rm K~km~s^{-1}$}}

\newcommand{\uSig}{\mbox{$\rm M_\odot~pc^{-2}$}}

\newcommand{\uSigSFR}{\mbox{$\rm M_\odot~yr^{-1}~kpc^{-2}$}}